\documentclass[pra,aps,epsfig,amsmath,amsbsy,showpacs]{revtex4}
\usepackage{amsmath,amsbsy}
\usepackage[dvips]{graphicx}
\begin{document}
\newcommand{\eps}{\ensuremath{\varepsilon}}
\newcommand{\mi}{\ensuremath{\mathrm{i}}}
\newcommand{\me}{\mathrm{e}}
\newcommand{\nn}{\nonumber \\}
\newcommand{\bra}[1]{\langle #1 |}
\newcommand{\ket}[1]{| #1 \rangle}
\newcommand{\bigbra}[1] {\big\langle #1\big|}
\newcommand{\bigket}[1] {\big|#1\big\rangle}
\newcommand{\Bigbra}[1] {\Big\langle #1\Big|}
\newcommand{\Bigket}[1] {\Big|#1\Big\rangle}
\newcommand{\dif}{\ensuremath{\mathrm{d}}}
\newcommand{\bx}{\boldsymbol{x}}
\newcommand{\br}{\boldsymbol{r}}
\newcommand{\dagg}{^{\dag}}
\newcommand{\balpha}{\ensuremath{\boldsymbol{\alpha}}}
\newcommand{\dd}[1]{\frac{\partial}{\partial #1}}
\newcommand{\Funcder}[1]{\frac{\delta }{\delta #1}}
\newcommand{\funcder}[2]{\frac{\delta #1}{\delta #2}}
\newcommand{\Partder}[1]{\frac{\partial }{\partial #1}}
\newcommand{\partder}[2]{\frac{\partial #1}{\partial #2}}
\newcommand{\Der}[1]{\frac{\dif}{\dif#1}}
\newcommand{\der}[2]{\frac{\dif#1}{\dif#2}}
\newcommand{\ave}[1]{\langle #1\rangle}
\newcommand{\Ave}[1]{\Big\langle #1\Big\rangle}
\newcommand{\intd}[1]{\int\frac{\dif #1}{2\pi}}
\newcommand{\tr}{\mathrm{tr}}
\newcommand{\half}{{\frac{1}{2}}}
\newcommand{\halfS}{{\textstyle\frac{1}{2}\,}}
\newcommand{\dint}{\int\!\!\!\int}
\newcommand{\intbr}{\int\dif\br\,}
\newcommand{\DelT}{{\Delta\mathrm{T}}}
\newcommand{\DelW}{{\Delta\mathrm{W}}}
\newcommand{\s}{\mathrm{s}}
\renewcommand{\H}{\hat{H}}
\newcommand{\T}{\hat{T}}
\newcommand{\W}{\hat{W}}
\newcommand{\U}{\hat{U}}
\renewcommand{\v}{\hat{v}}
\newcommand{\V}{\hat{V}}
\renewcommand{\t}{\hat{t}}
\newcommand{\dintbr}{\int\!\!\!\int\dif\br_1\dif\br_2\,}
\newcommand{\vsp}{\vspace{0.5cm}}
\newcommand{\Vsp}{\vspace{1cm}}
\newcommand{\Wsp}{\vspace{2cm}}
\newcommand{\hsp}{\hspace{0.5cm}}
\newcommand{\Hsp}{\hspace{1cm}}
\newcommand{\HHsp}{\hspace{2cm}}
\newcommand{\mhsp}{\hspace{-0.5cm}}
\newcommand{\mvsp}{\vspace{-0.5cm}}
\newcommand{\mvvsp}{\vspace{-0.25cm}}
\renewcommand{\tilde}{\widetilde}
\newcommand{\nlim}{n\rightarrow\infty}
\newcommand{\q}{$\quad$}
\newcommand{\qq}{$\qquad$}
\newcommand{\ö}{\"o}
\newcommand{\Ö}{\"O}
\newcommand{\ä}{\"a}
\newcommand{\rarr}{\rightarrow}
\newcommand{\lrarr}{\leftrightarrow}
\newcommand{\Rarr}{\Rightarrow}
\newcommand{\Lrarr}{\Longrightarrow}
\newcommand{\sumi}{\sum_{i=1}^N}
\newcommand{\sumj}{\sum_{j=1}^N}
\newcommand{\sumk}{\sum_{k=1}^n}
\newcommand{\suml}{\sum_{l=1}^n}
\newcommand{\textbs}[1]{\boldsymbol{#1}}
\newcommand{\bs}[1]{\boldsymbol{#1}}
\newcommand{\abs}[1]{|{#1}|}
\newcommand{\Abs}[1]{\big|{#1}\big|}
\renewcommand{\it}{\textit}
\newcommand{\bfit}[1]{{\it{#1}}}
\newcommand{\bR}{\cal{R}}
\newcommand{\norm}[1]{||{#1}||}
\newcommand{\Norm}[1]{\big|\big|{#1}\big|\big|}
\newcommand{\NORM}[1]{\Big|\Big|{#1}\Big|\Big|}
\renewcommand{\intbr}{\int\dif\br}
\renewcommand{\sp}[2]{\bra{#1}{#2}\rangle}
\newcommand{\SP}[2]{\Bigbra{#1}{#2}\Big\rangle}
\newcommand{\ul}{\underline}
\newcommand{\eq}{\eqref}
\newcommand{\Fr}{Fr\'echet }
\newcommand{\Ga}{G\^ateaux }
\newcommand{\epsn}{\ensuremath{\epsilon}}
\renewcommand{\rm}{\mathrm}
\renewcommand{\cal}{\mathcal}
\newcommand{\bb}[2]{\big\{{#1}\,\big|\:{#2}\big\}}
\newcommand{\BB}[2]{\Big\{{#1}\,\Big|\:{#2}\Big\}}
\newcommand{\LL}{L1\cap L3}
\renewcommand{\S}{\cal{S}}
\newcommand{\KS}{\mathrm{KS}}
\newcommand{\HF}{\mathrm{HF}}
\newcommand{\HFKS}{\mathrm{HFKS}}
\newcommand{\eff}{\mathrm{eff}}
\newcommand{\xc}{\mathrm{xc}}
\newcommand{\el}{\mathrm{el}}
\newcommand{\R}{\mathrm{R}}
\newcommand{\occ}{\mathrm{occ}}
\newcommand{\corr}{\mathrm{corr}}
\newcommand{\ext}{\mathrm{ext}}
\newcommand{\Coul}{\mathrm{Coul}}
\newcommand{\HK}{\mathrm{HK}}
\newcommand{\TF}{\mathrm{TF}}
\newcommand{\la}{\lambda}
\newcommand{\Dif}{\ensuremath{\mathrm{D}}}
\newcommand{\dr}{\delta\rho}
\newcommand{\exact}{\rm{exact}}
\renewcommand{\max}{\rm{max}}
\renewcommand{\min}{\rm{min}}



\newcommand{\LineH}[1]
{\linethickness{0.4mm}
\put(0,0){\line(1,0){#1}}
}
\newcommand{\LineS}[1]
{\linethickness{1mm}
\put(0,0){\line(1,0){#1}}
}

\newcommand{\LineWO}[1]
{\linethickness{0.5mm}
\put(0,0){\line(1,0){#1}}
}

\newcommand{\LineV}[1]
{\linethickness{0.4mm}
\put(0,0){\line(0,1){#1}}
}

\newcommand{\Linev}[1]
{\put(0.,0){\line(0,1){#1}}
}

\newcommand{\LineW}[1]
{\put(0.015,0){\line(0,1){#1}}
\put(0,0){\line(0,1){#1}}
\put(-0.015,0){\line(0,1){#1}}}

\newcommand{\LineHpt}[1]
{\put(0,0.015){\line(1,0){#1}}
\put(0,-0.015){\line(1,0){#1}}
\put(0,0){\circle*{0.15}}
\put(#1,0){\circle*{0.15}}}

\newcommand{\LineDl}[1]
{\put(0,0){\line(-1,-1){#1}}
\put(0.,0.01){\line(-1,-1){#1}}
\put(0,-0.01){\line(-1,-1){#1}}}

\newcommand{\Linedl}[1]
{\put(0.01,0.01){\line(-1,-2){#1}}
\put(-0.01,-0.01){\line(-1,-2){#1}}}

\newcommand{\LineUr}[1]
{\put(0,0.015){\line(1,1){#1}}
\put(0,0){\line(1,1){#1}}
\put(0,-0.015){\line(1,1){#1}}}

\newcommand{\LineDr}[1]
{\put(0,0.01){\line(1,-1){#1}}
\put(0,0){\line(1,-1){#1}}
\put(0,-0.01){\line(1,-1){#1}}}

\newcommand{\Linedr}[1]
{\put(0,0.01){\line(1,-3){#1}}
\put(0,0){\line(1,-3){#1}}
\put(0,-0.01){\line(1,-3){#1}}}

\newcommand{\Lineur}[1]
{\put(0,0.01){\line(1,3){#1}}
\put(0,0){\line(1,3){#1}}
\put(0,-0.01){\line(1,3){#1}}}

\newcommand{\DLine}[1]
{\put(0.,-0.05){\line(1,0){#1}}
\put(0.,0.05){\line(1,0){#1}}
\put(0,0){\circle*{0.1}}}

\newcommand{\Vector}[0]
{\thicklines\setlength{\unitlength}{1cm}\put(-0.13,0){\vector(-1,0){0}}}

\newcommand{\VectorR}[0]
{\thicklines\put(0.13,0.0){\vector(1,0){0}}}

\newcommand{\VectorUp}[0]
{\thicklines\setlength{\unitlength}{1cm}
\put(0.012,0.12){\vector(0,0){0}}
\put(-0.012,0.12){\vector(0,0){0}}}

\newcommand{\VectorDn}[0]
{\thicklines\setlength{\unitlength}{1cm}
\put(0.012,-0.12){\vector(0,-1){0}}
\put(-0.012,-0.12){\vector(0,-1){0}}}

\newcommand{\VectorDl}[0]
{\thicklines
\setlength{\unitlength}{1cm}
\put(-0.092,-0.076){\vector(-1,-1){0}}
\put(-0.076,-0.092){\vector(-1,-1){0}}
}

\newcommand{\VectorDr}[0]
{\thicklines
\setlength{\unitlength}{1cm}
\put(0.076,-0.092){\vector(1,-1){0}}
\put(0.092,-0.076){\vector(1,-1){0}}
}

\newcommand{\Vectordr}[0]
{\setlength{\unitlength}{1cm}
\put(0.022,0.112){\vector(1,-3){0}}
\put(0.052,0.118){\vector(1,-3){0}}
}

\newcommand{\Vectorur}[0]
{\setlength{\unitlength}{1cm}
\put(0.04,-0.062){\vector(1,3){0}}
\put(0.06,-0.068){\vector(1,3){0}}
}

\newcommand{\VectorUr}[0]
{\put(0.2,0.75){\vector(1,1){0}}
\put(0.18,0.69){\vector(1,1){0}}
\put(0.14,0.73){\vector(1,1){0}}
}

\newcommand{\VectorUl}[0]
{\put(-0.23,-0.02){\vector(-1,1){0}}
\put(-0.19,-0.03){\vector(-1,1){0}}
\put(-0.22,-0.06){\vector(-1,1){0}}
}

\newcommand{\Wector}[0]
{\put(-0.15,0)\Vector\put(0.15,0)\Vector}

\newcommand{\WectorUp}[0]
{\put(0,0.125)\VectorUp\put(0,-0.125)\VectorUp}

\newcommand{\WectorDn}[0]
{\put(0,0.125)\VectorDn\put(0,-0.125)\VectorDn}

\newcommand{\WectorDl}[0]
{\put(0.1,0.1)\VectorDl\put(-0.1,-0.1)\VectorDl}

\newcommand{\EllineH}[4]
{\put(0,0){\LineH{#1}}
\put(#2,0){\Vector}
\put(#2,0.45){\makebox(0,0){$#3$}}
\put(#2,-0.35){\makebox(0,0){$#4$}}}

\newcommand{\Elline}[4]
{\put(0,0){\LineV{#1}}
\put(0,#2){\VectorUp}
\put(-0.3,#2){\makebox(0,0){$#3$}}
\put(0.35,#2){\makebox(0,0){$#4$}}}

\newcommand{\DElline}[4]
{\put(0,0){\LineV{#1}}
\put(0,#2){\WectorUp}
\put(-0.3,#2){\makebox(0,0){$#3$}}
\put(0.3,#2){\makebox(0,0){$#4$}}}

\newcommand{\DEllineDn}[4]
{\put(0,0){\LineV{#1}}
\put(0,#2){\WectorDn}
\put(-0.25,#2){\makebox(0,0){$#3$}}
\put(0.25,#2){\makebox(0,0){$#4$}}}

\newcommand{\Ellinet}[4]
{\put(0,0){\Linev{#1}}
\put(0,#2){\VectorUp}
\put(-0.35,#2){\makebox(0,0){$#3$}}
\put(0.35,#2){\makebox(0,0){$#4$}}}

\newcommand{\EllineT}[4]
{\put(0,0){\LineW{#1}}
\put(0,#2){\VectorUp}
\put(-0.35,#2){\makebox(0,0){$#3$}}
\put(0.35,#2){\makebox(0,0){$#4$}}}

\newcommand{\EllineDnt}[4]
{\put(0,0){\Linev{#1}}
\put(0,#2){\VectorDn}
\put(-0.35,#2){\makebox(0,0){$#3$}}
\put(0.35,#2){\makebox(0,0){$#4$}}}

\newcommand{\EllineDn}[4]
{\put(0,0){\LineV{#1}}
\put(0,#2){\VectorDn}
\put(-0.35,#2){\makebox(0,0){$#3$}}
\put(0.35,#2){\makebox(0,0){$#4$}}}

\newcommand{\EllineDl}[4]
{\put(0,0){\LineDl{#1}}
\put(-#2,-#2){\VectorDl}
\put(-0.2,0.2){\makebox(-#1,-#1){$#3$}}
\put(0.2,-0.2){\makebox(-#1,-#1){$#4$}}}

\newcommand{\Ellinedl}[4]
{\put(0.01,0.01){\line(-1,-2){#1}}
\put(-0.01,-0.01){\line(-1,-2){#1}}
\thicklines\put(-0.05,-0.1){\vector(-1,-2){#2}}
\put(-0.2,0.2){\makebox(-#1,-#1){$#3$}}
\put(0.2,-0.2){\makebox(-#1,-#1){$#4$}}}

\newcommand{\EllineA}[7]
{\put(0.0,0.0){\line(#1,#2){#3}}
\put(0.005,0.0){\line(#1,#2){#3}}
\put(-0.005,0.0){\line(#1,#2){#3}}
\put(0,0){\vector(#1,#2){#4}}
\put(0.010,0){\vector(#1,#2){#4}}
\put(-0.010,0){\vector(#1,#2){#4}}
\put(#6,#7){\makebox(0,0){$#5$}}}

\newcommand{\EllineDr}[4]
{\put(0,0){\LineDr{#1}}
\put(#2,-#2){\VectorDr}
\put(#2,-#2){\makebox(-0.5,-0.5){$#3$}}
\put(#2,-#2){\makebox(0.5,0.5){$#4$}}}

\newcommand{\EllinedR}[5]
{\put(0,0){\line(1,-3){#1}}
\put(0.014,0){\line(1,-3){#1}}
\put(-0.014,0){\line(1,-3){#1}}
\put(#2,-#3){\makebox(0,0){{\Vectordr}}}
\put(#2,-#3){\makebox(-0.5,-0.5){$#4$}}
\put(#2,-#3){\makebox(0.5,0.5){$#5$}}}

\newcommand{\EllineuR}[5]
{\put(0,0){\line(1,3){#1}}
\put(0.014,0){\line(1,3){#1}}
\put(-0.014,0){\line(1,3){#1}}
\put(#2,#3){\makebox(0,0){\Vectorur}}
\put(#2,#3){\makebox(-0.5,-0.5){$#4$}}
\put(#2,#3){\makebox(0.5,0.5){$#5$}}}

\newcommand{\Ellinedr}[5]
{\put(0,0){\Linedr{#1}}
\put(#2,-#3){\makebox(0,0){\Vectordr}}
\put(#2,-#3){\makebox(-0.5,-0.5){$#4$}}
\put(#2,-#3){\makebox(0.5,0.5){$#5$}}}

\newcommand{\Ellineur}[5]
{\put(0,0){\Lineur{#1}}
\put(#2,#3){\makebox(0,0){\Vectorur}}
\put(#2,#3){\makebox(-0.5,-0.5){$#4$}}
\put(#2,#3){\makebox(0.5,0.5){$#5$}}}

\newcommand{\EllineUr}[4]
{\put(0,0){\LineUr{#1}}
\put(#2,#2){\makebox(-0.35,-0.35){\VectorUr}}
\put(-0.2,0.4){\makebox(#1,#1){$#3$}}
\put(0.2,-0.3){\makebox(#1,#1){$#4$}}}

\newcommand{\DEllineDl}[4]
{\put(0,0){\LineDl{#1}}
\put(-#2,-#2){\WectorDl}
\put(-0.25,0.25){\makebox(-#1,-#1){$#3$}}
\put(0.25,-0.25){\makebox(-#1,-#1){$#4$}}}

\newcommand{\Ebox}[2]
{\put(0,0){\LineH{#1}}
\put(0,#2){\LineH{#1}}
\put(0,0){\LineV{#2}}
\put(#1,0){\LineV{#2}}}

\newcommand{\dashH}
{\multiput(0.05,0)(0.25,0){5}{\line(1,0){0.15}}}

\newcommand{\dash}[1]
{\multiput(0.05,0)(0.25,0){#1}{\line(1,0){0.15}}}

\newcommand{\dashV}[1]
{\multiput(0.05,0)(0,0.25){#1}{\line(0,1){0.15}}}

\newcommand{\dashHp}
{\multiput(0.05,0)(0.25,0){6}{\line(1,0){0.15}}}

\newcommand{\DashH}
{\multiput(0.05,0)(0.25,0){10}{\line(1,0){0.15}}}

\newcommand{\dashHnum}[2]
{\multiput(0.05,0)(0.25,0){5}{\line(1,0){0.15}}
\put(-0.25,0){\makebox(0,0){$#1$}}
\put(1.5,0){\makebox(0,0){$#2$}}}

\newcommand{\dashHnuma}[2]
{\multiput(0.05,0)(0.25,0){5}{\line(1,0){0.15}}
\put(0.25,0.25){\makebox(0,0){$#1$}}
\put(1,0.25){\makebox(0,0){$#2$}}}

\newcommand{\dashHnumu}[2]
{\multiput(0.05,0)(0.25,0){5}{\line(1,0){0.15}}
\put(0.25,-0.25){\makebox(0,0){$#1$}}
\put(1,-0.25){\makebox(0,0){$#2$}}}

\newcommand{\Potint}
{\put(0,0)\dashH
\put(1.35,0){\makebox(0,0){x}}
\put(0,0){\circle*{0.15}}}

\newcommand{\potint}
{\multiput(0.05,0)(0.25,0){3}{\line(1,0){0.15}}
\put(0.85,0){\makebox(0,0){x}}
\put(0,0){\circle*{0.15}}}

\newcommand{\PotintS}
{\put(0,0){\dash{4}}
\put(1,0){\makebox(0,0){$\times$}}
\put(0,0){\circle*{0.15}}}

\newcommand{\PotintL}
{\put(-1.25,0)\dashH
\put(-1.35,0){\makebox(0,0){x}}
\put(0,0){\circle*{0.15}}}

\newcommand{\Effpot}
{\put(0,0)\dashH
\put(1.35,0){\makebox(0,0){x}}
\put(1.35,0){\circle{0.3}}
\put(0,0){\circle*{0.15}}}

\newcommand{\effpot}
{\multiput(0.05,0)(0.25,0){3}{\line(1,0){0.15}}
\put(0.85,0){\makebox(0,0){x}}
\put(0.85,0){\circle{0.3}}
\put(0,0){\circle*{0.1}}}

\newcommand{\EffpotL}
{\put(-1.25,0)\dashH
\put(0,0){\makebox(0,0){x}}
\put(-1.25,0){\circle{0.3}}
\put(-1.25,0){\circle*{0.15}}}

\newcommand{\Triang}
{\put(0,0){\line(2,1){0.5}}
\put(0,0){\line(2,-1){0.5}}
\put(0.5,-0.25){\line(0,1){0.5}}}

\newcommand{\TriangL}
{\put(0,0){\line(-2,1){0.5}}
\put(0,0){\line(-2,-1){0.5}}
\put(-0.5,-0.25){\line(0,1){0.5}}}

\newcommand{\hfint}
{\put(0,0)\dashH
\put(1.25,0){\makebox(0,0){\Triang}}
\put(0,0){\circle*{0.15}}}

\newcommand{\hfintL}
{\put(-1.25,0)\dashH
\put(-1.25,0){\makebox(0,0){\TriangL}}
\put(0,0){\circle*{0.15}}}

\newcommand{\VPloop}[1]
{\put(0,0){\circle{1}}
\put(0,0.0){\circle{1.}}
\put(0.02,0){\circle{1.}}
\put(0,0.02){\circle{1.}}
\put(0,-0.02){\circle{1}}
\put(-0.02,0){\circle{1}}
\put(0.52,0.05){\VectorDn}
\put(0.85,0){\makebox(0,0){$#1$}}}

\newcommand{\VPloopt}[1]
{\put(0,0){\circle{1}}
\put(0.52,0.05){\VectorDn}
\put(0.75,0){\makebox(0,0){$#1$}}}

\newcommand{\VPloopL}[1]
{\put(0,0){\circle{1}}
\put(0.01,0.){\circle{1}}\put(-0.01,0){\circle{1}}
\put(0,0.01){\circle{1}}\put(0,-0.01){\circle{1}}
\put(-0.5,0.05){\VectorDn}
\put(-0.75,0){\makebox(0,0){$#1$}}}

\newcommand{\VPloopLt}[1]
{\put(0,0){\circle{1}}
\put(-0.5,0){\VectorDn}
\put(-0.75,0){\makebox(0,0){$#1$}}}

\newcommand{\VPloopLR}[2]
{\put(0,0){\circle{1}}\put(-0.01,0){\circle{1}}
\put(0,0.01){\circle{1}}\put(0,-0.01){\circle{1}}
\put(-0.5,0){\VectorDn}
\put(0.5,0){\VectorUp}
\put(-0.75,0){\makebox(0,0){$#1$}}
\put(0.75,0){\makebox(0,0){$#2$}}}

\newcommand{\VPloopLRt}[2]
{\put(0,0){\circle{1}}
\put(-0.5,0){\VectorDn}
\put(0.5,0){\VectorUp}
\put(-0.75,0){\makebox(0,0){$#1$}}
\put(0.75,0){\makebox(0,0){$#2$}}}

\newcommand{\VPloopD}[2]
{\put(0,0){\circle{1}}
\put(0.01,0.){\circle{1}}\put(-0.01,0){\circle{1}}
\put(0.,0.){\circle{1}}\put(-0.01,0){\circle{1}}
\put(0,0.01){\circle{1}}\put(0,-0.01){\circle{1}}
\put(0,0.52){\VectorR}
\put(0,-0.52){\Vector}
\put(0,0.8){\makebox(0,0){$#1$}}
\put(0,-0.8){\makebox(0,0){$#2$}}}

\newcommand{\VPloopDt}[2]
{\put(0,0){\circle{1}}
\put(0,0.5){\VectorR}
\put(0.05,-0.47){\Vector}
\put(0,0.8){\makebox(0,0){$#1$}}
\put(0,-0.8){\makebox(0,0){$#2$}}}

\newcommand{\Loop}[2]
{\put(0,0){\oval(0.6,1.25)}\put(0.01,0.01){\oval(0.6,1.25)}\put(-0.01,-0.01){\oval(0.6,1.25)}
\put(0.3,0){\VectorUp}
\put(-0.3,0){\VectorDn}
\put(-0.65,0){\makebox(0,0){$#1$}}
\put(0.65,0){\makebox(0,0){$#2$}}}

\newcommand{\HFexch}[1]
{\put(0,0)\dashH
\qbezier(0,0.01)(0.625,0.515)(1.25,0.015)
\qbezier(0,-0.01)(0.625,0.485)(1.25,-0.015)
\put(0.625,0.26){\Vector}
\put(0.625,0.5){\makebox(0,0){$#1$}}
\put(0,0){\circle*{0.15}}
\put(1.25,0){\circle*{0.15}}}

\newcommand{\HFexcht}[1]
{\put(0,0)\dashH
\qbezier(0,0.01)(0.625,0.515)(1.25,0.015)
\put(0.625,0.26){\Vector}
\put(0.625,0.5){\makebox(0,0){$#1$}}
\put(0,0){\circle*{0.15}}
\put(1.25,0){\circle*{0.15}}}

\newcommand{\Dashpt}[1]
{\multiput(0,-0.6)(0,0.25){13}{\line(0,1){0.15}}
\put(0,0){\circle*{0.15}}
\put(0,#1){\circle*{0.15}}}

\newcommand{\photonH}[3]
{\qbezier(0,0)(0.08333,0.125)(0.1666667,0)
\qbezier(0.1666667,0)(0.25,-0.125)(0.3333333,0)
\qbezier(0.3333333,0)(0.416667,0.125)(0.5,0)
\qbezier(0.5,0)(0.583333,-0.125)(0.666667,0)
\qbezier(0.666667,0)(0.75,0.125)(0.833333,0)
\qbezier(0.833333,0)(0.916667,-0.125)(1,0)
\qbezier(1,0)(1.083333,0.125)(1.166667,0)
\qbezier(1.166667,0)(1.25,-0.125)(1.333333,0)
\qbezier(1.333333,0)(1.416667,0.125)(1.5,0)
\put(0.75,0.){\VectorR}
\put(0.75,0.35){\makebox(0,0){$#1$}}
\put(0,0){\circle*{0.15}}
\put(1.5,0){\circle*{0.15}}
\put(-0.5,0){\makebox(0,0){#2}}
\put(2,0){\makebox(0,0){#3}}}

\newcommand{\photon}[3]
{\qbezier(0,0)(0.08333,0.125)(0.1666667,0)
\qbezier(0.1666667,0)(0.25,-0.125)(0.3333333,0)
\qbezier(0.3333333,0)(0.416667,0.125)(0.5,0)

\qbezier(0.5,0)(0.583333,-0.125)(0.666667,0)
\qbezier(0.666667,0)(0.75,0.125)(0.833333,0)
\qbezier(0.833333,0)(0.916667,-0.125)(1,0)

\qbezier(1,0)(1.083333,0.125)(1.166667,0)
\qbezier(1.166667,0)(1.25,-0.125)(1.333333,0)
\qbezier(1.333333,0)(1.416667,0.125)(1.5,0)

\qbezier(1.5,0)(1.583333,-0.125)(1.666667,0)
\qbezier(1.666667,0)(1.75,0.125)(1.833333,0)
\qbezier(1.833333,0)(1.916667,-0.125)(2,0)
\put(1,0.0){\VectorR}
\put(1,0.35){\makebox(0,0){$#1$}}
\put(0,0){\circle*{0.15}}
\put(2,0){\circle*{0.15}}
\put(-0.35,0){\makebox(0,0){#2}}
\put(2.35,0){\makebox(0,0){#3}}}

\newcommand{\photonHS}[4]
{\qbezier(0,0)(0.08333,0.125)(0.1666667,0)
\qbezier(0.1666667,0)(0.25,-0.125)(0.3333333,0)
\qbezier(0.3333333,0)(0.416667,0.125)(0.5,0)
\qbezier(0.5,0)(0.583333,-0.125)(0.666667,0)
\qbezier(0.666667,0)(0.75,0.125)(0.833333,0)
\qbezier(0.833333,0)(0.916667,-0.125)(1,0)
\put(0.5,0.025){\VectorR}
\put(0.5,0.35){\makebox(0,0){$#1$}}
\put(0,0){\circle*{0.15}}
\put(1,0){\circle*{0.15}}
\put(0,-0.5){\makebox(0,0){#2}}
\put(1,-0.5){\makebox(0,0){#3}}}

\newcommand{\photonNE}[3]
{\qbezier(0,0)(0.22,-0.02)(0.2,0.2)
\qbezier(0.2,0.2)(0.18,0.42)(0.4,0.4)
\qbezier(0.4,0.4)(0.62,0.38)(0.6,0.6)
\qbezier(0.6,0.6)(0.58,0.82)(0.8,0.8)
\qbezier(0.8,0.8)(1.02,0.78)(1,1)
\qbezier(1,1)(0.98,1.22)(1.2,1.2)
\qbezier(1.2,1.2)(1.42,1.18)(1.4,1.4)
\qbezier(1.4,1.4)(1.38,1.62)(1.6,1.6)
\qbezier(1.6,1.6)(1.82,1.58)(1.8,1.8)
\qbezier(1.8,1.8)(1.78,2.02)(2,2)
\put(1,1){\makebox(0.05,-0.2){\VectorUp}}
\put(0,0){\circle*{0.15}}
\put(2,2){\circle*{0.15}}
\put(1,1){\makebox(-0.6,0.4){$#1$}}
\put(-0.35,-1){\makebox(0,2){$#2$}}
\put(2.35,1){\makebox(0,2){$#3$}}}

\newcommand{\photonNNE}[3]
{\qbezier(0,0)   (0.28,-0.02)(0.2,0.3)
\qbezier(0.2,0.3)(0.12,0.52)(0.4,0.6)
\qbezier(0.4,0.6)(0.68,0.58)(0.6,0.9)
\qbezier(0.6,0.9)(0.52,1.12)(0.8,1.2)
\qbezier(0.8,1.2)(1.08,1.18)(1,1.5)
\qbezier(1,1.5)  (0.92,1.72)(1.2,1.8)
\qbezier(1.2,1.8)(1.48,1.86)(1.4,2.1)
\qbezier(1.4,2.1)(1.365,2.24)(1.6,2.4)
\qbezier(1.6,2.4)(1.835,2.46)(1.8,2.7)
\qbezier(1.8,2.7)(1.765,2.84)(2,3)
\put(0.6,0.8){\makebox(0,0){\VectorUp}}
\put(0,0){\circle*{0.15}}
\put(2,3){\circle*{0.15}}
\put(1,0.8){\makebox(0,0){$#1$}}
\put(-0.35,0){\makebox(0,0){$#2$}}
\put(2.35,3){\makebox(0,0){$#3$}}}

\newcommand{\photonENE}[3]
{\qbezier(0,0)(0.17,-0.04)(0.2,0.1)
\qbezier(0.2,0.1)(0.23,0.32)(0.4,0.2)
\qbezier(0.4,0.2)(0.57,0.16)(0.6,0.3)
\qbezier(0.6,0.3)(0.63,0.52)(0.8,0.4)
\qbezier(0.8,0.4)(0.97,0.36)(1,0.5)
\qbezier(1,0.5)(1.03,0.72)(1.2,0.6)
\qbezier(1.2,0.6)(1.37,0.56)(1.4,0.7)
\qbezier(1.4,0.7)(1.43,0.92)(1.6,0.8)
\qbezier(1.6,0.8)(1.77,0.76)(1.8,0.9)
\qbezier(1.8,0.9)(1.83,1.12)(2,1)
\put(1.2,0.75){\makebox(-0.1,0.06){\VectorR}}
\put(0,0){\circle*{0.15}}
\put(2,1){\circle*{0.15}}
\put(1.2,1){\makebox(0,-0){$#1$}}
\put(-0.35,-1){\makebox(0,2){$#2$}}
\put(2.35,0){\makebox(0,2){$#3$}}}

\newcommand{\photonNW}[3]
{\qbezier(0,0)(-0.22,-0.02)(-0.2,0.2)
\qbezier(-0.2,0.2)(-0.18,0.42)(-0.4,0.4)
\qbezier(-0.4,0.4)(-0.62,0.38)(-0.6,0.6)
\qbezier(-0.6,0.6)(-0.58,0.82)(-0.8,0.8)
\qbezier(-0.8,0.8)(-1.02,0.78)(-1,1)
\qbezier(-1,1)    (-0.98,1.22)(-1.2,1.2)
\qbezier(-1.2,1.2)(-1.42,1.18)(-1.4,1.4)
\qbezier(-1.4,1.4)(-1.38,1.62)(-1.6,1.6)
\qbezier(-1.6,1.6)(-1.82,1.58)(-1.8,1.8)
\qbezier(-1.8,1.8)(-1.78,2.02)(-2,2)
\put(-1,1){\makebox(0,-0.2){\VectorUp}}
\put(0,0){\circle*{0.15}}
\put(-2,2){\circle*{0.15}}
\put(-1,1){\makebox(0.4,0.7){$#1$}}
\put(-2.35,2){\makebox(0,0){$#2$}}
\put(0.35,0){\makebox(0,0){$#3$}}}

\newcommand{\photonWNW}[3]
{\qbezier(0,0)(-0.17,-0.04)(-0.2,0.1)
\qbezier(-0.2,0.1)(-0.23,0.32)(-0.4,0.2)
\qbezier(-0.4,0.2)(-0.57,0.16)(-0.6,0.3)
\qbezier(-0.6,0.3)(-0.63,0.52)(-0.8,0.4)
\qbezier(-0.8,0.4)(-0.97,0.36)(-1,0.5)
\qbezier(-1,0.5)(-1.03,0.72)(-1.2,0.6)
\qbezier(-1.2,0.6)(-1.37,0.56)(-1.4,0.7)
\qbezier(-1.4,0.7)(-1.43,0.92)(-1.6,0.8)
\qbezier(-1.6,0.8)(-1.77,0.76)(-1.8,0.9)
\qbezier(-1.8,0.9)(-1.83,1.12)(-2,1)
\put(-1,1){\makebox(-0.2,-0.4){$\;$\VectorR}}
\put(0,0){\circle*{0.15}}
\put(-2,1){\circle*{0.15}}
\put(-1,1){\makebox(0.6,0.4){$#1$}}
\put(0.35,-1){\makebox(0,2){$#2$}}
\put(-2.35,1){\makebox(0,2){$#3$}}}

\newcommand{\Crossphotons}[6]
{\put(0,0){\photonNE{#5}{}{}}
\put(2,0){\photonNW{#6}{}{}}
\put(-0.35,0){\makebox(0,0){$#1$}}
\put(2.35,2){\makebox(0,0){$#2$}}
\put(2.35,0){\makebox(0,0){$#3$}}
\put(-0.35,2){\makebox(0,0){$#4$}}
}

\newcommand{\photonNe}[3]
{\qbezier(0,0)(0.22,-0.02)(0.2,0.2)
\qbezier(0.2,0.2)(0.18,0.42)(0.4,0.4)
\qbezier(0.4,0.4)(0.62,0.38)(0.6,0.6)
\qbezier(0.6,0.6)(0.58,0.82)(0.8,0.8)
\qbezier(0.8,0.8)(1.02,0.78)(1,1)
\qbezier(1,1)(0.98,1.22)(1.2,1.2)
\qbezier(1.2,1.2)(1.42,1.18)(1.4,1.4)
\qbezier(1.4,1.4)(1.38,1.5)(1.5,1.5)
\put(0.82,1.02){\makebox(0,0.01){\VectorR}}
\put(0,0){\circle*{0.15}}
\put(0.75,0.75){\makebox(-0.6,0.4){$#1$}}
\put(-0.35,0){\makebox(0,0){$#2$}}
\put(1.85,1.5){\makebox(0,0){$#3$}}}

\newcommand{\photonNw}[4]
{\qbezier(0,0)(0.02,0.22)(-0.2,0.2)
\qbezier(-0.2,0.2)(-0.42,0.18)(-0.4,0.4)
\qbezier(-0.4,0.4)(-0.38,0.62)(-0.6,0.6)
\qbezier(-0.6,0.6)(-0.82,0.58)(-0.8,0.8)
\qbezier(-0.8,0.8)(-0.78,1.02)(-1,1)
\qbezier(-1,1)    (-1.22,0.98)(-1.2,1.2)
\qbezier(-1.2,1.2)(-1.18,1.42)(-1.4,1.4)
\qbezier(-1.4,1.4)(-1.5,1.38)(-1.5,1.5)
\put(-#1,#1){\makebox(-0.1,-0.1){\VectorUp}}
\put(0,0){\circle*{0.15}}
\put(-1.5,1.5){\circle*{0.15}}
\put(-#1,#1){\makebox(1,0.3){$#2$}}
\put(0.35,0){\makebox(0,0){$#3$}}
\put(-1.85,1.5){\makebox(0,0){$#4$}}
}

\newcommand{\elstat}[3]
{\multiput(0.06,0)(0.25,0){8}{\line(1,0){0.15}}
\put(1,0.35){\makebox(0,0){$#1$}}
\put(0,0){\circle*{0.15}}
\put(2,0){\circle*{0.15}}
\put(-0.35,0){\makebox(0,0){#2}}
\put(2.35,0){\makebox(0,0){#3}}}

\newcommand{\elstatH}[3]
{\multiput(0.06,0)(0.25,0){6}{\line(1,0){0.15}}
\put(1,0.35){\makebox(0,0){$#1$}}
\put(0,0){\circle*{0.15}}
\put(1.5,0){\circle*{0.15}}
\put(-0.35,0){\makebox(0,0){#2}}
\put(2.35,0){\makebox(0,0){#3}}}

\newcommand{\elsta}[3]
{\multiput(0.06,0)(0.25,0){4}{\line(1,0){0.15}}
\put(1,0.35){\makebox(0,0){$#1$}}
\put(0,0){\circle*{0.1}}
\put(1,0){\circle*{0.1}}
\put(-0.35,0){\makebox(0,0){#2}}
\put(1.35,0){\makebox(0,0){#3}}}

\newcommand{\elstatNO}[3]
{\multiput(0.06,0.08)(0.01,0.01){14}{\tiny.}
\multiput(0.30,0.32)(0.01,0.01){14}{\tiny.}
\multiput(0.55,0.57)(0.01,0.01){14}{\tiny.}
\multiput(0.79,0.81)(0.01,0.01){14}{\tiny.}
\multiput(1.03,1.05)(0.01,0.01){14}{\tiny.}
\multiput(1.27,1.29)(0.01,0.01){14}{\tiny.}
\multiput(1.51,1.53)(0.01,0.01){14}{\tiny.}
\multiput(1.75,1.77)(0.01,0.01){14}{\tiny.}
\put(0.95,0.75){\makebox(0,0){\VectorUr}}
\put(0,0){\circle*{0.15}}
\put(2,2){\circle*{0.15}}
\put(0.75,1.25){\makebox(0,0){$#1$}}
\put(-0.5,0){\makebox(0,0){$#2$}}
\put(2.5,2){\makebox(0,0){$#3$}}
}

\newcommand{\elstatNW}[3]
{\multiput(-0.05,0.05)(-0.015,0.015){10}{\circle*{0.02}}
\multiput(-0.3,0.3)(-0.015,0.015){10}{\circle*{0.02}}
\multiput(-0.55,0.55)(-0.015,0.015){10}{\circle*{0.02}}
\multiput(-0.8,0.8)(-0.015,0.015){10}{\circle*{0.03}}
\multiput(-1.05,1.05)(-0.015,0.015){10}{\circle*{0.03}}
\multiput(-1.3,1.3)(-0.015,0.015){10}{\circle*{0.03}}
\multiput(-1.55,1.55)(-0.015,0.015){10}{\circle*{0.03}}
\multiput(-1.8,1.8)(-0.015,0.015){10}{\circle*{0.03}}
\put(-0.9,0.9){\makebox(0,0){\VectorUl}}
\put(0,0){\circle*{0.215}}
\put(-2,2){\circle*{0.2153}}
\put(-0.9,0.9){\makebox(0,0){\VectorUl}}
\put(0,0){\circle*{0.215}}
\put(-2,2){\circle*{0.215}}
\put(-0.5,1){\makebox(0,0){$#1$}}
\put(0,-0.5){\makebox(0,0){$#2$}}
\put(-2,2.5){\makebox(0,0){$#3$}}
}

\newcommand{\photonSE}[5]
{\put(0,0){\photonHS{#3}{#4}{}{}}
\put(1.5,0){\VPloopD{#1}{#2}}
\put(2,0){\photonHS{#5}{}{}{}}}

\newcommand{\photonSEt}[5]
{\put(0,0){\photonHS{#3}{#4}{}{}}
\put(1.5,0){\VPloopDt{#1}{#2}}
\put(2,0){\photonHS{#5}{}{}{}}}

\newcommand{\ElSE}[3]
{\qbezier(0,-1)(.2025,-1.1489)(0.3420,-0.9397)
\qbezier(0.3420,-0.9397)(0.4167,-0.7217)(0.6428,-0.766)
\qbezier(0.6428,-0.766)(0.8937,-0.7499)(0.866,-0.5)
\qbezier(0.866,-0.5)(0.7831,-0.2850)(0.9848,-0.1736)
\qbezier(0.9848,-0.1736)(1.1667,0)(0.9848,0.1736)
\qbezier(0.9848,0.1736)(0.7831,0.2850)(0.866,0.5)
\qbezier(0.866,0.5)(0.8937,0.7499)(0.6428,0.766)
\qbezier(0.6428,0.766)(0.4167,0.7217)(0.3420,0.9397)
\qbezier(0.3420,0.9397)(.2025,1.1489)(0,1)
\put(1.05,0.02){\VectorUp}
\put(0,1){\circle*{0.15}}
\put(0,-1){\circle*{0.15}}
\put(1.45,0){\makebox(0,0){$#1$}}
\put(-0.35,-1){\makebox(0,0){#2}}
\put(-0.35,1){\makebox(0,0){#3}}}

\newcommand{\ElSEL}[3]
{\qbezier(0,-1)(-.2025,-1.1489)(-0.3420,-0.9397)
\qbezier(-0.3420,-0.9397)(-0.4167,-0.7217)(-0.6428,-0.766)
\qbezier(-0.6428,-0.766)(-0.8937,-0.7499)(-0.866,-0.5)
\qbezier(-0.866,-0.5)(-0.7831,-0.2850)(-0.9848,-0.1736)
\qbezier(-0.9848,-0.1736)(-1.1667,0)(-0.9848,0.1736)
\qbezier(-0.9848,0.1736)(-0.7831,0.2850)(-0.866,0.5)
\qbezier(-0.866,0.5)(-0.8937,0.7499)(-0.6428,0.766)
\qbezier(-0.6428,0.766)(-0.4167,0.7217)(-0.3420,0.9397)
\qbezier(-0.3420,0.9397)(-.2025,1.1489)(0,1)
\put(-1.05,0.02){\VectorUp}
\put(0,1){\circle*{0.15}}
\put(0,-1){\circle*{0.15}}
\put(-1.45,0){\makebox(0,0){$#1$}}
\put(0.35,-1){\makebox(0,0){#2}}
\put(0.35,1){\makebox(0,0){#3}}}

\newcommand{\SEpolt}[5]
{\qbezier(0,-1.5)(.2025,-1.6489)(0.3420,-1.4397)
\qbezier(0.3420,-1.4397)(0.4167,-1.2217)(0.6428,-1.266)
\qbezier(0.6428,-1.266)(0.8937,-1.2499)(0.866,-1)
\qbezier(0.866,-1)(0.7831,-0.7850)(0.9848,-0.6736)
\qbezier(1,-0.5)(1.1,-0.5)(0.9848,-0.6736)
\qbezier(1,0.5)(1.1,0.5)(0.9848,0.6736)
\qbezier(0.9848,0.6736)(0.7831,0.7850)(0.866,1)
\qbezier(0.866,1)(0.8937,1.2499)(0.6428,1.266)
\qbezier(0.6428,1.266)(0.4167,1.2217)(0.3420,1.4397)
\qbezier(0.3420,1.4397)(.2025,1.6489)(0,1.5)
\put(1,0){\VPloopLRt{#1}{#2}}
\put(0.87,-1){\VectorUp}
\put(0.67,1.23){\Vector}
\put(1.3,-1){\makebox(0,0){#3}}
\put(1,0.5){\circle*{0.15}}
\put(1,-0.5){\circle*{0.15}}
\put(0,1.5){\circle*{0.15}}
\put(0,-1.5){\circle*{0.15}}
\put(-0.35,-1.5){\makebox(0,0){#4}}
\put(-0.35,1.5){\makebox(0,0){#5}}}

\newcommand{\SEpoltNA}[5]
{\qbezier(0,-1.5)(.2025,-1.6489)(0.3420,-1.4397)
\qbezier(0.3420,-1.4397)(0.4167,-1.2217)(0.6428,-1.266)
\qbezier(0.6428,-1.266)(0.8937,-1.2499)(0.866,-1)
\qbezier(0.866,-1)(0.7831,-0.7850)(0.9848,-0.6736)
\qbezier(1,-0.5)(1.1,-0.5)(0.9848,-0.6736)
\qbezier(1,0.5)(1.1,0.5)(0.9848,0.6736)
\qbezier(0.9848,0.6736)(0.7831,0.7850)(0.866,1)
\qbezier(0.866,1)(0.8937,1.2499)(0.6428,1.266)
\qbezier(0.6428,1.266)(0.4167,1.2217)(0.3420,1.4397)
\qbezier(0.3420,1.4397)(.2025,1.6489)(0,1.5)
\put(1,0){\circle{1}}
\put(0.87,-1){\VectorUp}
\put(0.67,1.23){\Vector}
\put(1.3,-1){\makebox(0,0){#3}}
\put(1,0.5){\circle*{0.15}}
\put(1,-0.5){\circle*{0.15}}
\put(0,1.5){\circle*{0.15}}
\put(0,-1.5){\circle*{0.15}}
\put(-0.35,-1.5){\makebox(0,0){#4}}
\put(-0.35,1.5){\makebox(0,0){#5}}}

\newcommand{\SEpol}[5]
{\qbezier(0,-1.5)(.2025,-1.6489)(0.3420,-1.4397)
\qbezier(0.3420,-1.4397)(0.4167,-1.2217)(0.6428,-1.266)
\qbezier(0.6428,-1.266)(0.8937,-1.2499)(0.866,-1)
\qbezier(0.866,-1)(0.7831,-0.7850)(0.9848,-0.6736)
\qbezier(1,-0.5)(1.1,-0.5)(0.9848,-0.6736)
\qbezier(1,0.5)(1.1,0.5)(0.9848,0.6736)
\qbezier(0.9848,0.6736)(0.7831,0.7850)(0.866,1)
\qbezier(0.866,1)(0.8937,1.2499)(0.6428,1.266)
\qbezier(0.6428,1.266)(0.4167,1.2217)(0.3420,1.4397)
\qbezier(0.3420,1.4397)(.2025,1.6489)(0,1.5)
\put(1,0){\VPloopLR{#1}{#2}}
\put(0.87,-1){\VectorUp}
\put(0.67,1.23){\Vector}
\put(1.3,-1){\makebox(0,0){#3}}
\put(1,0.5){\circle*{0.15}}
\put(1,-0.5){\circle*{0.15}}
\put(0,1.5){\circle*{0.15}}
\put(0,-1.5){\circle*{0.15}}
\put(-0.35,-1.5){\makebox(0,0){#4}}
\put(-0.35,1.5){\makebox(0,0){#5}}}

\setlength{\unitlength}{1cm} \thicklines

\title{Construction of accurate Kohn-Sham potentials for the lowest states
of the helium atom: Accurate test of the ionization-potential
theorem}
\author{I. Lindgren\footnote{ingvar.lindgren@fy.chalmers.se}, S.
Salomonson\footnote{f3asos@fy.chalmers.se}, and F.
M\"oller\footnote{gu99frmo@dd.chalmers.se}}
\affiliation{Department of Physics, Chalmers University of
Technology and the G\"oteborg University,\\ G\"oteborg,
Sweden}

\begin{abstract}
Accurate local Kohn-Sham potentials have been constructed for the
ground $1s^2\,^1S$ state and, in particular, for the lowest
triplet $1s2s\,^{3}S$ state of the helium atom, using electron
densities from many-body calculations and the procedure of van
Leeuwen and Baerends (Phys. Rev. A{\bf49}, 2138 (1994)). The
resulting Kohn-Sham orbitals reproduce the many-body densities
very accurately, and furthermore we have demonstrated that the
negative of the energy eigenvalue of the outermost electron
orbital agrees with the corresponding ionization energy with
extreme accuracy. The procedure is also applied to the
Hartree-Fock density of the $1s2s\,^{3}S$ state, and the Kohn-Sham
eigenvalue of the $2s$ orbital is found to agree very well with
the corresponding Hartree-Fock eigenvalue, which is the negative
of the ionization energy in this model due to Koopmans' theorem.
The results for the $1s2s\,^{3}S$ state clearly demonstrate that
there is no conflict between the locality of the Kohn-Sham
potential and the exclusion principle, as claimed by Nesbet (Phys.
Rev. A{\bf58}, R12 (1998)).
\end{abstract}


 \maketitle

\normalsize \section{The Kohn-Sham model}

According to the Hohenberg-Kohn (HK)
theorem~\cite{HK64,PY89,DG90}, the energy of any electronic system
can be expressed as a functional of the electron density,
$\rho(\br)$,
\begin{equation}
  \label{EHK}
  E[\rho]=F_\HK[\rho]+\int\dif\br\,\rho(\br)v(\br),
\end{equation}
where $v(\br)$ is the external potential and $F_\HK[\rho]$ is the
universal HK functional, which in the constrained-search
formulation is~\cite{Le79,Li83}
\begin{equation}
  \label{FHK}
  F_\HK[\rho]=\min_{\Psi\rarr\rho}\bigbra{\Psi}\T+
  \W\bigket{\Psi}.
\end{equation}
Here, $\T$ is the kinetic-energy and $\W$ the
electron-electron-interaction operators of the system (in atomic
units),
\begin{equation}
  \label{TW}
  \T=-\sum_{i=1}^N\halfS\nabla^2_i\,; \quad   \W=\sum_{i<j}^N
\frac{1}{|\br_i-\br_j|}.
\end{equation}
The wave function, $\Psi$, is normalized and belongs to the
\it{Sobolev space} $H^1(\bR^{3N})$~\cite{Li83,Lee03}, and the
corresponding functional is defined for all $N$-representable
densities~\cite{DG90}. The ground-state energy of the system is
obtained by minimizing the energy functional over these
densities~\cite{DG90},
\begin{equation}
  \label{MinEHK}
  E_{0}=\min_{\rho\rarr N}E[\rho]=E[\rho_0].
\end{equation}
This leads to the Euler-Lagrange equation
\begin{equation}
  \label{ELHK}
  \funcder{F_\HK[\rho]}{\rho(\br)}+v(\br)=\mu,
\end{equation}
where $\mu$ is the Lagrange parameter for the normalization
constraint, $\intbr\,\rho(\br)=N$.

In the Kohn-Sham (KS) model the interacting system is replaced by
a system of \it{noninteracting} electrons, moving in the local KS
potential, $v_\KS(\br)$~\cite{KS65},
\begin{equation}
  \label{KS}
\big[-\halfS\nabla^2+v_\KS(\br)\big]\phi_i(\br)=\eps_{i}\,\phi_i(\br).
\end{equation}
The energy functional for this system is
\begin{equation}
  \label{EKS}
  E_\KS[\rho]=T_\KS[\rho]+\int\dif\br\,\rho(\br)\,v_\KS(\br),
\end{equation}
where
\begin{equation}
  \label{rho}
   \rho(\br)=\sum_{i=1}^N |\phi_i(\br)|^2
\end{equation}
is the electron density. The kinetic-energy functional is
\begin{equation}
  \label{TKS}
   T_\KS[\rho]=\min_{\Phi\rarr\rho}\bigbra{\Phi}\T
  \bigket{\Phi},
\end{equation}
where $\Phi=\det\{\phi_i\}$ is a single Slater-determinantal wave
function. Minimizing this functional leads to the Euler-Lagrange
equation
\begin{equation}
  \label{MinKE}
  \funcder{T_\KS[\rho]}{\rho(\br)}+v_\KS(\br)=\mu.
\end{equation}
Comparing with Eq. \eq{ELHK}, leads -- apart from an additive
constant -- to the relation
\begin{equation}
  \label{VKS}
  v_\KS(\br)=\funcder{F_\HK[\rho]}{\rho(\br)}-\funcder{T_\KS[\rho]}{\rho(\br)}
  +v(\br).
\end{equation}

The Hohenberg-Kohn-Sham model was originally proven for the ground
state but it was demonstrated by Gunnarsson and
Lundqvist~\cite{GL76} that it is valid also for the lowest state
of a given symmetry. Later it has been shown to hold also for more
general excited states~\cite{Goerl99,LevyNagy99}.

Although the form of the KS potential is generally not known, it
can be constructed with arbitrary accuracy in cases where the
electron density is known from other sources, e.g., from
experiments or from \it{ab initio} calculations. Essentially two
schemes have been developed for this purpose, by Zhao and
Parr~\cite{ZParr93,WParr93,ZMParr94} and by van Leeuwen and
Baerends~\cite{vLB94}, respectively.

The KS orbitals were originally assumed to have no other physical
significance than generating the exact electron density, but it
was later found by Perdew et al.~\cite{PPL82,LPS84,PL97} and
independently by Almbladh and Pendroza~\cite{AP84} that the
eigenvalue of the outermost electron (with opposite sign) equals
the ionization energy of the system. Perdew et al. have shown that
considering densities that integrate to non-integrals,
\begin{equation}
  \label{N}
  M=\intbr\,\rho(\br),
\end{equation}
the theorem holds in the range $N-1<M<N$ and hence when this
number approaches $N$ from below. This condition is known as the
\it{ionization-potential theorem}~\cite{PL97}.

The validity of the ionization-potential theorem has been
challenged by Kleinman~\cite{Kl97} with counterarguments supplied
by Perdew and Levy~\cite{PL97}. A number of numerical
verifications of the theorem have been performed in the
past~\cite{ZP93,vLB94,GLB95,LG96,MKR97,NL98,CGB02,ZND03,Harb04},
generally with low or moderate accuracy due to problems in
representing the density accurately using an analytical basis
set~\cite{ZMParr94,MKR97}. In the present work we use a numerical
basis set, which has made it possible to demonstrate the validity
of the theorem with much higher accuracy than in any previous
calculation known to us~\footnote{A short version of this work has
been submitted for publication in Phys. Rev. Letters}.

The construction of the KS potential from the electron density has
so far mainly been performed for atomic and molecular ground
states. Recently, however, Harbola~\cite{Harb04} has constructed
the potential for the first excited singlet state, $1s2s\,^1S$, of
the helium atom.

Our primary goal for the present work has been to construct the KS
potential for an excited \it{triplet} state, which, as far as we
know, has not been done before. It has been rigorously shown that
the KS potential is under general conditions strictly
\it{local}~\cite{EE84,EE84a,Li83,Lee03}, and this has also been
demonstrated in some of our previous
works~\cite{LS03a,LS03b,LS04}. Nevertheless, this fact has been
disputed in several papers by Nesbet~\cite{Ne98,Ne01,Ne03}, who
claims that the locality condition is in conflict with the
exclusion principle. In the helium triplet state the two electrons
can have the same spin orientation, and therefore our result
represents a final rebuttal of the objection of Nesbet.

\section{Construction of the Kohn-Sham potential from electron density}

In the present work we apply the scheme of van Leeuwen and
Baerends to construct accurate KS potentials for the lowest states
of the helium atom. Following van Leeuwen and Baerends, we obtain
after multiplying the KS equations \eq{KS} from the left by
$\phi_i^*(\br)$ and summing over the $N$ electrons
\begin{equation}
  \label{vKS}
  v_\KS(\br)\,\rho(\br)=\sum_{i=1}^N\Big[\halfS\phi_i^*(\br)\nabla^2\phi_i(\br)
  +\eps_i\Abs{\phi_i(\br)}^2\Big],
\end{equation}
where $\rho(\br)$ is the electron density
\begin{equation}
  \label{dens}
 \rho(\br)=\sum_{i=1}^N\Abs{\phi_i(\br)}^2.
\end{equation}
This leads to a self-consistency problem, which can be solved by
iteration. Defining the electronic part of the potential,
$v_\el(\br)$, by
\begin{equation}
  \label{vel}
  v_\KS(\br)=-\frac{Z}{r}+v_\el(\br)+ \rm{const.},
\end{equation}
the solution is obtained by means of the formula
\begin{equation}
  \label{iter}
  v_\el^{k+1}(\br)=\frac{\rho_k(\br)}{\rho_0(\br)}\,v_\el^k(\br),
\end{equation}
where $\rho_0(\br)$ is the exact many-body density and
$\rho_k(\br)$ is the density generated with the potential
$v_\el^k(\br)$. This procedure is continued until certain
convergence criteria are met.

\section{Many-Body Theory}
The many-body electron density needed for this procedure has been
evaluated by means of many-body perturbation
technique~\cite{LM86}, using the nonrelativistic pair-correlation
program developed by Salomonson and \Öster~\cite{SO89a}. We shall
briefly indicate this procedure here.

\begin{figure}
\begin{center}
\begin{picture}(4,4)(-4,-1)
\put(-2,1){\Elline{1.5}{0.75}{}{}}
\put(-2,-0.5){\DElline{1.5}{0.75}{}{}}
\put(-1,1){\Elline{1.5}{0.75}{}{}}
\put(-1,-0.5){\DElline{1.5}{0.75}{}{}} \put(-2,1){\LineS{1}}
\put(0,1){\makebox(0,0){\LARGE =}}
\end{picture}
\begin{picture}(2.5,3)(-3,-1)
\put(-2,1){\Elline{1.5}{0.75}{}{}}
\put(-2,-0.5){\DElline{1.5}{0.75}{}{}}
\put(-1,1){\Elline{1.5}{0.75}{}{}}
\put(-1,-0.5){\DElline{1.5}{0.75}{}{}} \put(-2,1){\dash{4}}
\end{picture}
\begin{picture}(3,3)(-1,-1)
\put(-0.75,1){\makebox(0,0){\LARGE +}}
\put(0,1.5){\Elline{1}{0.5}{}{}} \put(0,0.5){\Elline{1}{0.5}{}{}}
\put(0,-0.5){\DElline{1}{0.5}{}{}}
\put(1,1.5){\Elline{1}{0.5}{}{}} \put(1,0.5){\Elline{1}{0.5}{}{}}
\put(1,-0.5){\DElline{1}{0.5}{}{}} \put(0,1.5){\dash{4}}
\put(0,0.5){\LineS{1}}
\end{picture}
\begin{picture}(3,3)(-0.5,-1)
\put(-0.75,1){\makebox(0,0){\Huge -}}
\put(0,1.75){\Elline{1}{0.5}{}{}}
\put(0,0.75){\DElline{1}{0.5}{}{}}
\put(0,-0.75){\DElline{1}{0.5}{}{}}
\put(1,1.75){\Elline{1}{0.5}{}{}}
\put(1,0.75){\DElline{1}{0.5}{}{}}
\put(1,-0.75){\DElline{1}{0.5}{}{}} \put(0,1.75){\LineS{1}}
\put(-0.25,0.75){\LineWO{1.5}} \put(-0.25,0.25){\LineWO{1.5}}
\put(-0.25,0.25){\LineV{0.5}}\put(1.25,0.25){\LineV{0.5}}
\end{picture}\\
\begin{picture}(3,3.5)(-1.5,-1)
\put(-2,0.5){\makebox(0,0){\LARGE$\Delta E$\;=}}
\put(0,0.75){\DElline{1}{0.5}{}{}}
\put(0,-0.75){\DElline{1}{0.5}{}{}}
\put(1,0.75){\DElline{1}{0.5}{}{}}
\put(1,-0.75){\DElline{1}{0.5}{}{}} \put(-0.25,0.75){\LineWO{1.5}}
\put(-0.25,0.25){\LineWO{1.5}}
\put(-0.25,0.25){\LineV{0.5}}\put(1.25,0.25){\LineV{0.5}}
\end{picture}
\renewcommand{\normalsize}{\small}
\caption{Upper line: Graphical representation of the pair equation
(Eq. \eq{Pair}). The vertical lines represent the valence orbitals
(double arrows) and virtual orbitals (single arrow). The thick
horizontal line represents $\Omega_2$, the dotted line the
electrostatic interaction between the electrons and the box the
effective two-body interaction $W_2$ \eq{Eff2}. Lower line:
Graphical representation of the energy shift due to the
perturbation (Eq. \eq{E}).}
  \label{Fig:Omega2}
\end{center}
\end{figure}
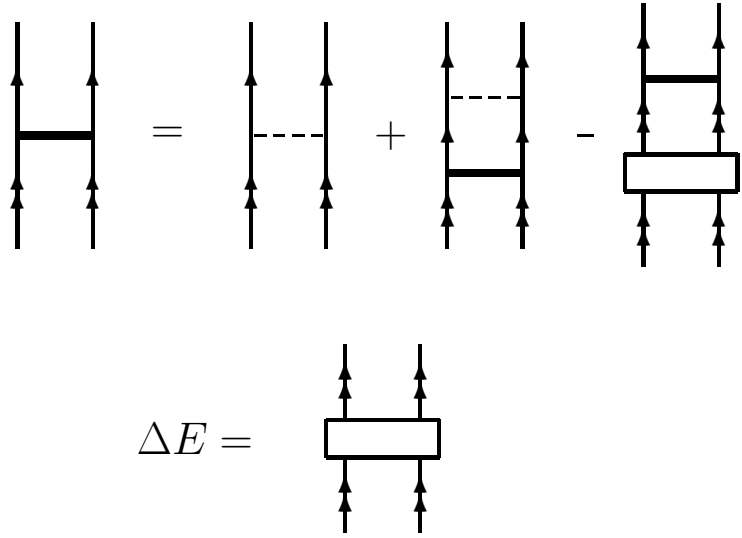
We want to solve the Schr\ödinger equation
\begin{equation}
  \label{Schreq}
  H\,\Psi=E\,\Psi
\end{equation}
and partition the Hamiltonian into a zeroth-order hamiltonian and
a perturbation
\begin{equation}
  \label{Part}
  H=H_0+H'.
\end{equation}
We start from a zeroth-order or \it{model function} $\Psi_0$,
which is an eigenfunction of $H_0$,
\begin{equation}
  \label{Modelf}
  H_0\,\Psi_0=E_0\,\Psi_0.
\end{equation}
The exact solution can be expressed
\begin{equation}
  \label{Waveop}
  \Psi=\Omega\,\Psi_0,
\end{equation}
where $\Omega$ is the \it{wave operator}, satisfying the
\it{generalized Bloch equation} in the \it{linked-diagram form}
 \begin{equation}
  \label{Bloch}
   \big[\Omega,H_0\big]P=\big(H'\Omega-\Omega
   W\big)_{\rm{linked}}P.
\end{equation}
Here, $W$ is the \it{effective interaction}, in intermediate
normalization ($\Psi_0=P\Psi$) given by $W=PH'\Omega P$. $P$ is
the projection operator for the model space, which in this simple
case is assumed to contain only a single model state, $\Psi_0$.
Only so-called linked diagrams will contribute according to the
linked-diagram theorem~\cite{LM86}.

Using second quantization, the wave operator can be separated into
normal-ordered one-, two-,.. body parts
\begin{equation}
  \label{Omega1}
  \Omega=1+ \Omega_1+\Omega_2+\dots
\end{equation}
or
\begin{equation}
  \label{WOSQ}
  \Omega=1+\{a\dagg_ia_j\}x_j^i+\frac{1}{2!}\,\{a\dagg_ia\dagg_ja_la_k\}\,x_{kl}^{ij}
  +\cdots
\end{equation}
using the sum convention. $a\dagg/a$ are the electron
creation/annihilation operators, and the curly brackets denote the
normal-ordering. The $n$-body part of the wave operator then
satisfies the equation
\begin{equation}
  \label{CCA}
   \big[\Omega_n,H_0\big]P=\big(H'\Omega-\Omega W\big)_{\rm{linked},n}P.
\end{equation}

\section{Application to the lowest states of the helium atom}

For heliumlike systems, starting from hydrogenlike orbitals, the
wave operator can be expressed by means of the two-body part only
of the wave operator,
\begin{equation}
  \label{Omega2}
  \Omega=1+ \Omega_2,
\end{equation}
satisfying the 'pair equation'
\begin{equation}
  \label{Pair}
   \big[\Omega_2,H_0\big]P=\big(H'\Omega-\Omega\, W_2\big)_{\rm{linked},2}P.
\end{equation}
This equation is exhibited graphically in the upper part of Fig.
\ref{Fig:Omega2}. Here, the thick line represents $\Omega_2$ and
the box the two-body part of the effective interaction
\begin{equation}
  \label{Eff2}
  W_2=\big(PH'\Omega P\big)_2.
\end{equation}

The total energy of the system is
\begin{equation}
  \label{E}
  E=E_0+\Delta E,
\end{equation}
where the energy shift, $\Delta E$, is in this case given by
\begin{equation}
  \label{DeltaE}
  \Delta E=\bra{\Psi_0}H'\Omega\ket{\Psi_0}=\bra{\Psi_0}W_2\ket{\Psi_0}
\end{equation}
and represented graphically by all 'closed' two-body diagrams, as
indicated at the bottom part of Fig. \ref{Fig:Omega2}. Since the
final state of the ionization process is in our cases the ground
state of the He$^+$ ion, with the exact nonrelativistic energy of
-2 H, the binding energy of the outermost electron becomes (in
atomic units)
\begin{equation}
  \label{BE}
  \rm{BE}=-2-E.
\end{equation}

\begin{figure}[htbp]
  \begin{center}
 \includegraphics[width=8cm]{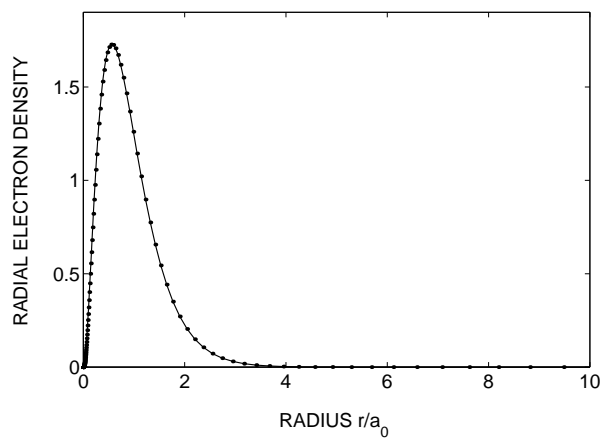}
    \caption{The Kohn-Sham density (dots) superimposed on the many-body
    density (solid line) for the helium ground state.}
 \label{nks1s2}
  \end{center}
\end{figure}

\begin{figure}[htbp]
  \begin{center}
 \includegraphics[width=8cm]{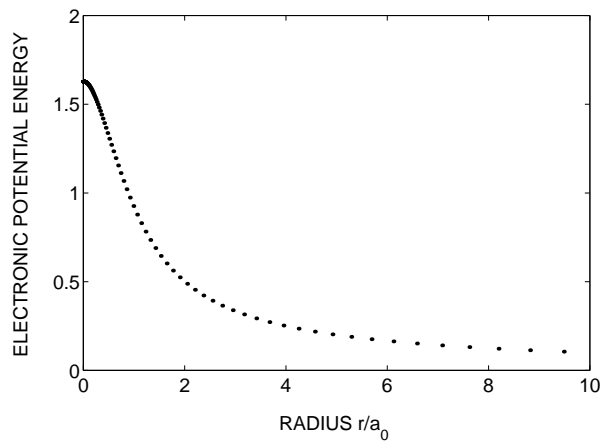}
    \caption{The electronic part of the Kohn-Sham potential for the helium
    ground state.}
    \label{vks1s2}
  \end{center}
\end{figure}

In the present work the pair equation \eq{Pair} has been solved,
using the numerical procedure developed by Salomonson and
\Öster~\cite{SO89a}, and densities for the $1s^2\,^1S$ ground
state and the lowest triplet state, $1s2s\,^{3}S$, of the helium
atom have been evaluated. These densities are then used to
construct the corresponding KS potentials, as discussed above. The
wave functions obtained in this way are virtually exact, apart
from relativistic, mass-polarization and quantum-electrodynamical
effects. In a similar fashion we have also used the Hartree-Fock
density to construct the corresponding KS potential.

It can be argued that the procedure used here corresponds to
approaching the electron-density integral \eq{N} to the electron
number from below, $M\rarr N-0$~\cite{PL97}, and hence the
ionization-potential theorem can be tested.

In order to achieve good accuracy, particularly for the eigenvalue
of the outermost electron orbital, it is important to have the
exact density in an accurate form and to have this density well
reproduced by the Kohn-Sham orbitals. In the present work we have
generated the many-body density using a large numerical grid, and
the convergence criteria are set so that the Kohn-Sham density
should not deviate from the many-body density by more than one
part in $10^9$ at any point. The convergence rate was usually
quite slow, and several thousands of iterations were often needed
to reach this level of accuracy. To improve the convergence rate
and avoid 'oscillations', it was sometimes helpful to take some
average of the last two iterations as the input for the next one.
It is also important to keep the electronic part, $v_\el$, of the
KS-potential positive at all points by adjusting the constant in
Eq. \eq{vel}. After the iteration procedure was completed, the
constant is determined so that the potential approaches zero as
$r\rarr\infty$.

\begin{figure}[htbp]
  \begin{center}
 \includegraphics[width=8cm]{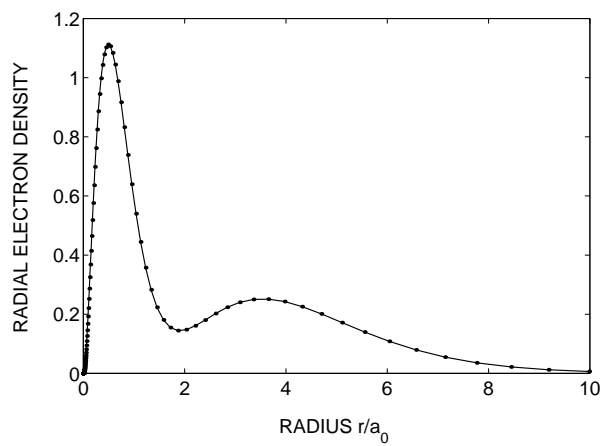}
    \caption{The Kohn-Sham density (dots) superimposed on the many-body
    density (solid line) for $1s2s^3S$.}
 \label{nks}
  \end{center}
\end{figure}

\begin{figure}[htbp]
  \begin{center}
 \includegraphics[width=8cm]{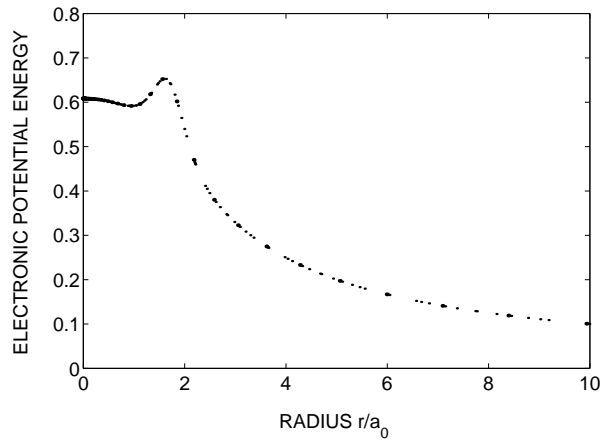}
    \caption{The electronic part of the Kohn-Sham potential for the $1s2s^3S$
    state of helium.}
    \label{vks}
  \end{center}
\end{figure}

As in our earlier works~\cite{SO89a}, we apply an exponential
radial grid $r_i=e^{x_i}/Z$, where ${x_i}$ is a discrete linear
lattice with equally spaced points with $x$ typically ranging from
$x_{\min}=-11$ to about $x_{\max}=4$. For the triplet state at
least four different grids were used, and the results extrapolated
in the standard way. Also the end point of the grid, $x_\max$, is
varied, as it was found that the Kohn-Sham results were quite
sensitive to that value (See Fig. \ref{xmaxextr}), most likely due
to the fact that our pair functions are forced to be zero at the
end point. This has very little effect on standard many-body
calculations, but the KS eigenvalue depends strongly upon the tail
of the density distribution and therefore more affected by the
boundary condition. Hence, an extrapolation of $x_{\max}$ is
required in the KS case. It is likely that the results could be
improved by more sophisticated boundary conditions.

In the evaluation of the many-body electron density a partial-wave
expansion is used~\cite{SO89a}, normally up to $l_\max=10$, and an
extrapolation performed in the standard way. In our procedure,
however, the Kohn-Sham potential is evaluated for successive
truncations of the partial-wave expansion, and the above-mentioned
ionization-potential theorem could be tested for each truncation
separately, as well as after the $l_\max$ extrapolation.

\subsection{The many-body density of the helium ground state}

As a preliminary test of our procedure, we have applied this to
the ground state of the helium atom, where the Kohn-Sham potential
has previously been constructed~\cite{MKR97,Harb04}. The electron
density obtained from the KS orbitals is shown in Fig.
\ref{nks1s2} (dots) superimposed on the many-body density (dots).
In this figure, the two densities are indistinguishable. The
resulting KS potential is shown in Fig. \ref{vks1s2}.

These calculations have been performed for a single grid with 201
points, with $x_\max=3.6$ and $l_\max=10$ without any
extrapolations. The results obtained is then -0.903 7041 H for the
KS $1s$ eigenvalue and 0.903 7052 H for the many-body ionization
energy, which verifies the above-mentioned ionization-potential
theorem to 5-6 digits. By careful extrapolations the
pair-correlation approach yields the value 0.903 724 39
H~\cite{SO89a}, which agrees well with the very accurate value
obtained by Frankowski and Pekeris~\cite{FP66} and by Freund et
al.~\cite{FHM84} of 0.903 724 377 H (uncorrected for relativity,
mass-polarization and QED effects).

In the corresponding calculation by Harbola~\cite{Harb04}, an
electron density taken from the literature~\cite{Koga93} was used.
Two different configurations, $1s^2\,^1S$ and $1s2s\,^1S$,
respectively, were used, and the energy eigenvalue of the highest
occupied orbital was in both cases found to be 0.899 H.

\subsection{The many-body density of the lowest triplet state of helium}

\begin{table}[htbp]
     \begin{center}
\caption{Comparison between the Kohn-Sham $2s$ eigenvalue and the
many-body ionization energy for the $1s2s\,^3S$ state of helium
with $l_\max=10$ and different end points of the numerical grid.}
    \begin{tabular}{|c|c|c|}\\
\hline
      $x_\max$ & KS eigenvalue   & Many-body IP\\\hline
       3.8 & -0.175 228 7967  & 0.175 229 3578\\
       4.0 & -0.175 229 1111 & 0.175 229 3634\\
       4.2 & -0.175 229 2488 & 0.175 229 3649\\
       4.4 & -0.175 229 3135 & 0.175 229 3634\\\hline
  extrapol & -0.175 229 3630 & 0.175 229 3639\\\hline
    \end{tabular}
     \label{Tab:extrapol}
 \end{center}
\end{table}

\begin{table}[htbp]
   \begin{center}
\caption{Comparison between the Kohn-Sham $2s$ eigenvalue and the
many-body ionization energy for the $1s2s\,^3S$ state of helium
with different truncations of the partial-wave expansion.}
    \begin{tabular}{|c|c|c|}\\
\hline
      $l_\max$ & KS eigenvalue   & Many-body IP\\\hline
       4 & -0.175 228 6206 & 0.175 228 6214\\
       6 & -0.175 229 2341 & 0.175 229 2354\\
       8 & -0.175 229 3366 & 0.175 229 3379\\
      10 & -0.175 229 3630 & 0.175 229 3639\\\hline
  extrapol & -0.175 229 3794 & 0.175 229 3797\\\hline
    \end{tabular}
       \label{Tab:extrapol2}
 \end{center}
\end{table}

\begin{figure}[htbp]
  \begin{center}
 \includegraphics[width=8cm]{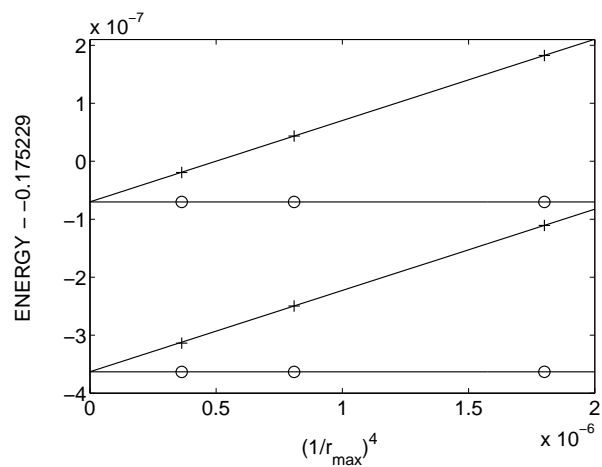}
    \caption{$x_\max$ extrapolation of the $2s$ eigenvalue (rings)
    and the negative ionization energy (crosses) for the
    $1s2s\,^3S$ state of helium with $l_\max=5$ and 10.}
 \label{xmaxextr}
  \end{center}
\end{figure}

\begin{figure}[htbp]
  \begin{center}
 \includegraphics[width=8cm]{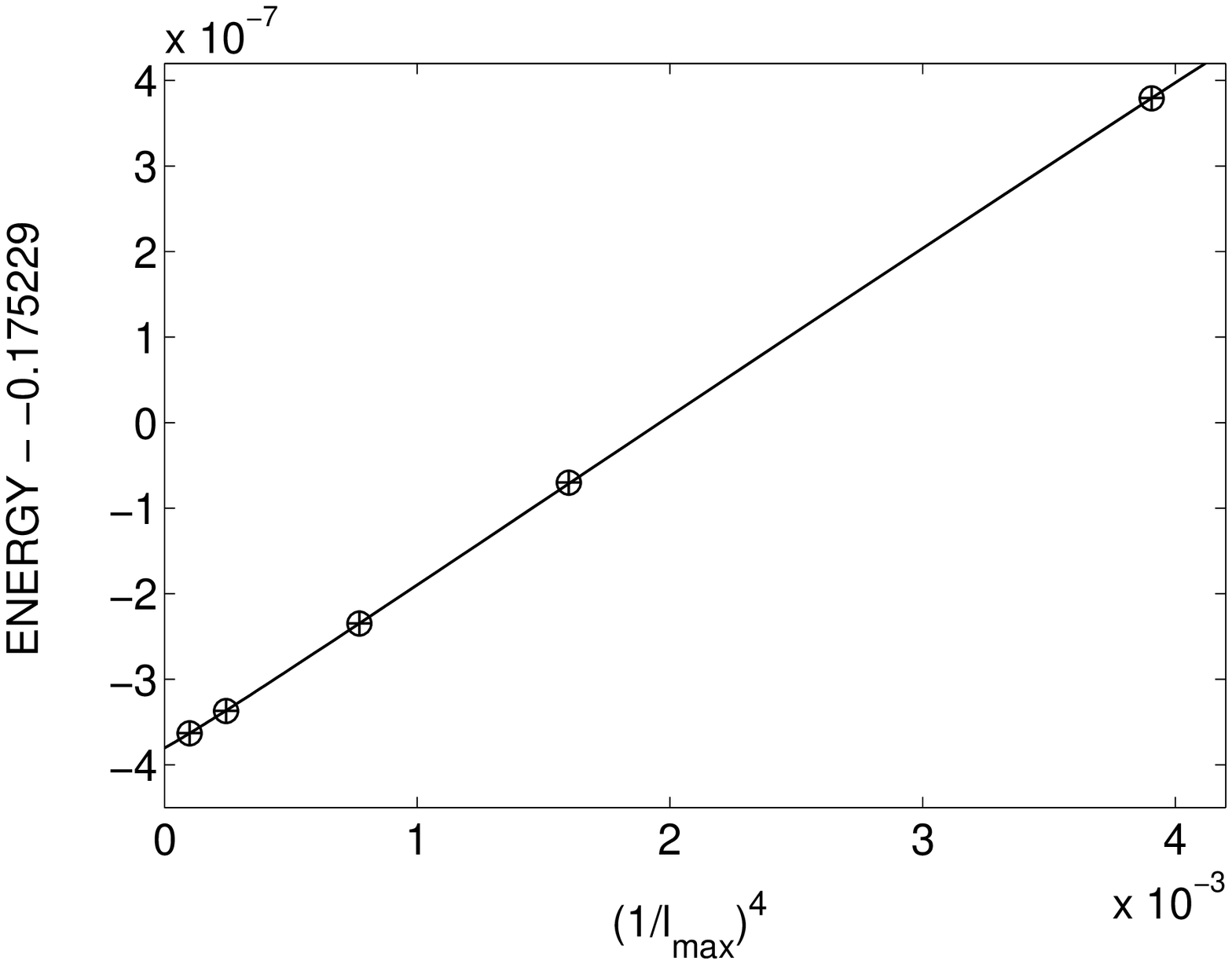}
    \caption{$l_\max$ extrapolation of the $2s$ eigenvalue and the
    negative ionization energy for
    $1s2s^3S$ of helium.}
 \label{lmaxextr}
  \end{center}
\end{figure}

As mentioned, our primary goal of the present work has been to
construct the Kohn-Sham potential from the many-body density for
the lowest triplet state, $1s2s\;^3S$, of the helium atom, and in
this case we have performed extensive extrapolations, as we shall
demonstrate below.

The final KS-density for the $1s2s^3S$ system is shown in Fig.
\ref{nks} (dots), together with the many-body density (solid
line). The corresponding KS-potential is shown in Fig. \ref{vks}.
It is interesting to note that the potential has a 'bump' close to
the node of the outermost (valence) electron. This is typical of
this kind of potential and is an effect of the electron \it{self
interaction} (SIC)~\cite{Li71,TS76,PZ81}. This depends
approximately on $\rho_{\rm{val}}^{1/3}$, where $\rho_{\rm{val}}$
is the density of the valence electron, and hence varies strongly
near the node of the valence orbital.

In Table \ref{Tab:extrapol} we show the KS $2s$ eigenvalue and the
corresponding many-body ionization energy after grid extrapolation
with the partial-wave expansion truncated at $l_\max=10$ and
different values of the grid end point, $x_\max$. This
extrapolation is illustrated graphically for two values of
$l_\max$ in Fig. \ref{xmaxextr}. It is found that the KS
eigenvalue is -- in contrast to the many-body ionization energy --
quite sensitive to the end point. After the $x_\max$
extrapolation, the values are found to agree to nine digits for
each partial wave truncation.

In order to find the "true" values of these quantities, it is
necessary also to extrapolate the partial-wave expansion, with the
result shown in Table \ref{Tab:extrapol2}. This is illustrated in
Fig. \ref{lmaxextr}. These final values represents the
nonrelativistic ionization energy, as before uncorrected for mass
polarization as well as for relativistic and QED effects. These
values agree to eight digits with the corresponding value 0,175
229 3782 H, obtained by Pekeris~\cite{Pek59}.

\subsection{The Hartree-Fock density of the lowest triplet state of
helium}

As an additional test of the procedure described here, we have
applied this also to the Hartree-Fock density of lowest triplet
state of the helium atom. This density is generated by solving the
standard HF equations and then inserted into the generating
formula (\ref{iter}) in place of the many-body density. The
resulting densities are quite similar to those given above, as is
the resulting KS potential, since HF is quite a good approximation
for this system.

In the HF approximation the orbital eigenvalues correspond exactly
to the corresponding ionization energies (with opposite sign), and
therefore a comparison of the Hartree-Fock-Kohn-Sham (HFKS) $2s$
eigenvalue with the corresponding HF value would constitute a
further test of the above-mentioned ionization-potential theorem.
Here, we found that the agreement is extremely good without any
$x_\max$ extrapolation. As an illustration we give the values
obtained after grid extrapolation for $x_\max=4$, where the HFKS
$2s$ eigenvalue is -0.174 256 072 542 H and the HF value -0.174
256 072 544 H -- an agreement to 11 digits! After complete
extrapolations the result is -0.174 256 0724 H, which is expected
to be accurate to 8-9 digits.

\section{Summary and comments}
We have demonstrated that it is possible to construct a local
Kohn-Sham potential for the lowest triplet state of the helium
atom with extreme accuracy. The agreement between the absolute
values of the ionization potential and highest-lying KS orbital
energy eigenvalue is verified to nine digits, which -- as far as
we know -- represents by far the most accurate numerical test of
the ionization-potential theorem performed to date. This result
also clearly demonstrates that there is no conflict between the
locality theorem and the exclusion principle, as claimed by
Nesbet~\cite{Ne98,Ne01,Ne03}.

\bibliographystyle{prsty}


\begin{thebibliography}{44}
\expandafter\ifx\csname natexlab\endcsname\relax\def\natexlab#1{#1}\fi
\expandafter\ifx\csname bibnamefont\endcsname\relax
  \def\bibnamefont#1{#1}\fi
\expandafter\ifx\csname bibfnamefont\endcsname\relax
  \def\bibfnamefont#1{#1}\fi
\expandafter\ifx\csname citenamefont\endcsname\relax
  \def\citenamefont#1{#1}\fi
\expandafter\ifx\csname url\endcsname\relax
  \def\url#1{\texttt{#1}}\fi
\expandafter\ifx\csname urlprefix\endcsname\relax\def\urlprefix{URL }\fi
\providecommand{\bibinfo}[2]{#2}
\providecommand{\eprint}[2][]{\url{#2}}

\bibitem[{\citenamefont{Hohenberg and Kohn}(1964)}]{HK64}
\bibinfo{author}{\bibfnamefont{P.}~\bibnamefont{Hohenberg}} \bibnamefont{and}
  \bibinfo{author}{\bibfnamefont{W.}~\bibnamefont{Kohn}},
  \bibinfo{journal}{Phys. Rev.} \textbf{\bibinfo{volume}{136}},
  \bibinfo{pages}{B864} (\bibinfo{year}{1964}).

\bibitem[{\citenamefont{Parr and Yang}(1989)}]{PY89}
\bibinfo{author}{\bibfnamefont{R.~G.} \bibnamefont{Parr}} \bibnamefont{and}
  \bibinfo{author}{\bibfnamefont{W.}~\bibnamefont{Yang}},
  \emph{\bibinfo{title}{Density-Functional Theory of Atoms and Molecules}}
  (\bibinfo{publisher}{Oxford Univ. Press}, \bibinfo{address}{New York},
  \bibinfo{year}{1989}).

\bibitem[{\citenamefont{Dreizler and Gross}(1990)}]{DG90}
\bibinfo{author}{\bibfnamefont{R.~M.} \bibnamefont{Dreizler}} \bibnamefont{and}
  \bibinfo{author}{\bibfnamefont{E.~K.~U.} \bibnamefont{Gross}},
  \emph{\bibinfo{title}{\textit{Density Functional Theory}}}
  (\bibinfo{publisher}{Springer-Verlag}, \bibinfo{address}{Berlin},
  \bibinfo{year}{1990}).

\bibitem[{\citenamefont{Levy}(1979)}]{Le79}
\bibinfo{author}{\bibfnamefont{M.}~\bibnamefont{Levy}}, \bibinfo{journal}{Proc.
  Natl. Acad. Sci. USA} \textbf{\bibinfo{volume}{76}}, \bibinfo{pages}{6062}
  (\bibinfo{year}{1979}).

\bibitem[{\citenamefont{Lieb}(1983)}]{Li83}
\bibinfo{author}{\bibfnamefont{E.~H.} \bibnamefont{Lieb}},
  \bibinfo{journal}{Int. J. Quantum Chem.} \textbf{\bibinfo{volume}{24}},
  \bibinfo{pages}{243} (\bibinfo{year}{1983}).

\bibitem[{\citenamefont{van Leeuwen}(2003)}]{Lee03}
\bibinfo{author}{\bibfnamefont{R.}~\bibnamefont{van Leeuwen}},
  \bibinfo{journal}{Adv. Quantum Chem.} \textbf{\bibinfo{volume}{43}},
  \bibinfo{pages}{25} (\bibinfo{year}{2003}).

\bibitem[{\citenamefont{Kohn and Sham}(1965)}]{KS65}
\bibinfo{author}{\bibfnamefont{W.}~\bibnamefont{Kohn}} \bibnamefont{and}
  \bibinfo{author}{\bibfnamefont{L.~J.} \bibnamefont{Sham}},
  \bibinfo{journal}{Phys. Rev.} \textbf{\bibinfo{volume}{140}},
  \bibinfo{pages}{A1133} (\bibinfo{year}{1965}).

\bibitem[{\citenamefont{Gunnarsson and Lundqvist}(1976)}]{GL76}
\bibinfo{author}{\bibfnamefont{O.}~\bibnamefont{Gunnarsson}} \bibnamefont{and}
  \bibinfo{author}{\bibfnamefont{B.~I.} \bibnamefont{Lundqvist}},
  \bibinfo{journal}{Phys. Rev. B} \textbf{\bibinfo{volume}{10}},
  \bibinfo{pages}{4274} (\bibinfo{year}{1976}).

\bibitem[{\citenamefont{G{\"o}rling}(1999)}]{Goerl99}
\bibinfo{author}{\bibfnamefont{A.}~\bibnamefont{G{\"o}rling}},
  \bibinfo{journal}{Phys. Rev. A} \textbf{\bibinfo{volume}{59}},
  \bibinfo{pages}{3359} (\bibinfo{year}{1999}).

\bibitem[{\citenamefont{Levy and Nagy}(1999)}]{LevyNagy99}
\bibinfo{author}{\bibfnamefont{M.}~\bibnamefont{Levy}} \bibnamefont{and}
  \bibinfo{author}{\bibfnamefont{A.}~\bibnamefont{Nagy}},
  \bibinfo{journal}{Phys. Rev. Lett.} \textbf{\bibinfo{volume}{83}},
  \bibinfo{pages}{4361} (\bibinfo{year}{1999}).

\bibitem[{\citenamefont{Zhao and Parr}(1993{\natexlab{a}})}]{ZParr93}
\bibinfo{author}{\bibfnamefont{Q.}~\bibnamefont{Zhao}} \bibnamefont{and}
  \bibinfo{author}{\bibfnamefont{R.~G.} \bibnamefont{Parr}},
  \bibinfo{journal}{J. Chem. Phys.} \textbf{\bibinfo{volume}{98}},
  \bibinfo{pages}{543} (\bibinfo{year}{1993}{\natexlab{a}}).

\bibitem[{\citenamefont{Wang and Parr}(1993)}]{WParr93}
\bibinfo{author}{\bibfnamefont{Y.}~\bibnamefont{Wang}} \bibnamefont{and}
  \bibinfo{author}{\bibfnamefont{R.~G.} \bibnamefont{Parr}},
  \bibinfo{journal}{Phys. Rev. A} \textbf{\bibinfo{volume}{47}},
  \bibinfo{pages}{R1591} (\bibinfo{year}{1993}).

\bibitem[{\citenamefont{Zhao et~al.}(1994)\citenamefont{Zhao, Morrison, and
  Parr}}]{ZMParr94}
\bibinfo{author}{\bibfnamefont{Q.}~\bibnamefont{Zhao}},
  \bibinfo{author}{\bibfnamefont{R.~C.} \bibnamefont{Morrison}},
  \bibnamefont{and} \bibinfo{author}{\bibfnamefont{R.~G.} \bibnamefont{Parr}},
  \bibinfo{journal}{Phys. Rev. A} \textbf{\bibinfo{volume}{50}},
  \bibinfo{pages}{2138} (\bibinfo{year}{1994}).

\bibitem[{\citenamefont{van Leeuwen and Baerends}(1994)}]{vLB94}
\bibinfo{author}{\bibfnamefont{R.}~\bibnamefont{van Leeuwen}} \bibnamefont{and}
  \bibinfo{author}{\bibfnamefont{E.~J.} \bibnamefont{Baerends}},
  \bibinfo{journal}{Phys. Rev. A} \textbf{\bibinfo{volume}{49}},
  \bibinfo{pages}{2421} (\bibinfo{year}{1994}).

\bibitem[{\citenamefont{Perdew et~al.}(1982)\citenamefont{Perdew, Parr, Levy,
  and Baldus.Jr.}}]{PPL82}
\bibinfo{author}{\bibfnamefont{J.~P.} \bibnamefont{Perdew}},
  \bibinfo{author}{\bibfnamefont{R.~G.} \bibnamefont{Parr}},
  \bibinfo{author}{\bibfnamefont{M.}~\bibnamefont{Levy}}, \bibnamefont{and}
  \bibinfo{author}{\bibfnamefont{J.~L.} \bibnamefont{Baldus.Jr.}},
  \bibinfo{journal}{Phys. Rev. Lett.} \textbf{\bibinfo{volume}{49}},
  \bibinfo{pages}{1691} (\bibinfo{year}{1982}).

\bibitem[{\citenamefont{Levy et~al.}(1984)\citenamefont{Levy, Perdew, and
  Sahni}}]{LPS84}
\bibinfo{author}{\bibfnamefont{M.}~\bibnamefont{Levy}},
  \bibinfo{author}{\bibfnamefont{J.~P.} \bibnamefont{Perdew}},
  \bibnamefont{and} \bibinfo{author}{\bibfnamefont{V.}~\bibnamefont{Sahni}},
  \bibinfo{journal}{Phys. Rev. A} \textbf{\bibinfo{volume}{30}},
  \bibinfo{pages}{2745} (\bibinfo{year}{1984}).

\bibitem[{\citenamefont{Perdew and Levy}(1997)}]{PL97}
\bibinfo{author}{\bibfnamefont{J.~P.} \bibnamefont{Perdew}} \bibnamefont{and}
  \bibinfo{author}{\bibfnamefont{M.}~\bibnamefont{Levy}},
  \bibinfo{journal}{Phys. Rev. B} \textbf{\bibinfo{volume}{56}},
  \bibinfo{pages}{16021} (\bibinfo{year}{1997}).

\bibitem[{\citenamefont{Almbladh and Pedroza}(1984)}]{AP84}
\bibinfo{author}{\bibfnamefont{C.-O.} \bibnamefont{Almbladh}} \bibnamefont{and}
  \bibinfo{author}{\bibfnamefont{A.~C.} \bibnamefont{Pedroza}},
  \bibinfo{journal}{Phys. Rev. A} \textbf{\bibinfo{volume}{29}},
  \bibinfo{pages}{2322} (\bibinfo{year}{1984}).

\bibitem[{\citenamefont{Kleinman}(1997)}]{Kl97}
\bibinfo{author}{\bibfnamefont{L.}~\bibnamefont{Kleinman}},
  \bibinfo{journal}{Phys. Rev. B} \textbf{\bibinfo{volume}{56}},
  \bibinfo{pages}{12042} (\bibinfo{year}{1997}).

\bibitem[{\citenamefont{Zhao and Parr}(1993{\natexlab{b}})}]{ZP93}
\bibinfo{author}{\bibfnamefont{Q.}~\bibnamefont{Zhao}} \bibnamefont{and}
  \bibinfo{author}{\bibfnamefont{R.~G.} \bibnamefont{Parr}},
  \bibinfo{journal}{J. Chem. Phys.} \textbf{\bibinfo{volume}{98}},
  \bibinfo{pages}{543} (\bibinfo{year}{1993}{\natexlab{b}}).

\bibitem[{\citenamefont{Gritsenko et~al.}(1995)\citenamefont{Gritsenko, van
  Leeuwen, and Baerends}}]{GLB95}
\bibinfo{author}{\bibfnamefont{O.~V.} \bibnamefont{Gritsenko}},
  \bibinfo{author}{\bibfnamefont{R.}~\bibnamefont{van Leeuwen}},
  \bibnamefont{and} \bibinfo{author}{\bibfnamefont{E.~J.}
  \bibnamefont{Baerends}}, \bibinfo{journal}{Phys. Rev. A}
  \textbf{\bibinfo{volume}{52}}, \bibinfo{pages}{1870} (\bibinfo{year}{1995}).

\bibitem[{\citenamefont{Levy and G{\"o}rling}(1996)}]{LG96}
\bibinfo{author}{\bibfnamefont{M.}~\bibnamefont{Levy}} \bibnamefont{and}
  \bibinfo{author}{\bibfnamefont{A.}~\bibnamefont{G{\"o}rling}},
  \bibinfo{journal}{Phys. Rev. B} \textbf{\bibinfo{volume}{53}},
  \bibinfo{pages}{969} (\bibinfo{year}{1996}).

\bibitem[{\citenamefont{Mura et~al.}(1997)\citenamefont{Mura, Knowles, and
  Reynolds}}]{MKR97}
\bibinfo{author}{\bibfnamefont{M.~E.} \bibnamefont{Mura}},
  \bibinfo{author}{\bibfnamefont{P.~J.} \bibnamefont{Knowles}},
  \bibnamefont{and} \bibinfo{author}{\bibfnamefont{C.~A.}
  \bibnamefont{Reynolds}}, \bibinfo{journal}{J. Chem. Phys.}
  \textbf{\bibinfo{volume}{106}}, \bibinfo{pages}{9659} (\bibinfo{year}{1997}).

\bibitem[{\citenamefont{Nagy and Levy}(1998)}]{NL98}
\bibinfo{author}{\bibfnamefont{{\'A}.}~\bibnamefont{Nagy}} \bibnamefont{and}
  \bibinfo{author}{\bibfnamefont{M.}~\bibnamefont{Levy}},
  \bibinfo{journal}{Chem. Phys. Letters} \textbf{\bibinfo{volume}{296}},
  \bibinfo{pages}{313} (\bibinfo{year}{1998}).

\bibitem[{\citenamefont{Chong et~al.}(2002)\citenamefont{Chong, Gritsenko, and
  Baerends}}]{CGB02}
\bibinfo{author}{\bibfnamefont{D.~P.} \bibnamefont{Chong}},
  \bibinfo{author}{\bibfnamefont{O.~V.} \bibnamefont{Gritsenko}},
  \bibnamefont{and} \bibinfo{author}{\bibfnamefont{E.~J.}
  \bibnamefont{Baerends}}, \bibinfo{journal}{J. Phys. Chem.}
  \textbf{\bibinfo{volume}{116}}, \bibinfo{pages}{1760} (\bibinfo{year}{2002}).

\bibitem[{\citenamefont{Zhan et~al.}(2003)\citenamefont{Zhan, Nichols, and
  Dixon}}]{ZND03}
\bibinfo{author}{\bibfnamefont{C.-G.} \bibnamefont{Zhan}},
  \bibinfo{author}{\bibfnamefont{J.~A.} \bibnamefont{Nichols}},
  \bibnamefont{and} \bibinfo{author}{\bibfnamefont{D.}~\bibnamefont{Dixon}},
  \bibinfo{journal}{J. Phys. Chem. A} \textbf{\bibinfo{volume}{107}},
  \bibinfo{pages}{4184} (\bibinfo{year}{2003}).

\bibitem[{\citenamefont{Harbola}(2004)}]{Harb04}
\bibinfo{author}{\bibfnamefont{M.~K.} \bibnamefont{Harbola}},
  \bibinfo{journal}{Phys. Rev. A} \textbf{\bibinfo{volume}{69}},
  \bibinfo{pages}{042512} (\bibinfo{year}{2004}).

\bibitem[{\citenamefont{Englisch and Englisch}(1984{\natexlab{a}})}]{EE84}
\bibinfo{author}{\bibfnamefont{H.}~\bibnamefont{Englisch}} \bibnamefont{and}
  \bibinfo{author}{\bibfnamefont{R.}~\bibnamefont{Englisch}},
  \bibinfo{journal}{Phys. Stat. Sol.} \textbf{\bibinfo{volume}{123}},
  \bibinfo{pages}{711} (\bibinfo{year}{1984}{\natexlab{a}}).

\bibitem[{\citenamefont{Englisch and Englisch}(1984{\natexlab{b}})}]{EE84a}
\bibinfo{author}{\bibfnamefont{H.}~\bibnamefont{Englisch}} \bibnamefont{and}
  \bibinfo{author}{\bibfnamefont{R.}~\bibnamefont{Englisch}},
  \bibinfo{journal}{Phys. Stat. Sol.} \textbf{\bibinfo{volume}{124}},
  \bibinfo{pages}{373} (\bibinfo{year}{1984}{\natexlab{b}}).

\bibitem[{\citenamefont{Lindgren and Salomonson}(2003{\natexlab{a}})}]{LS03a}
\bibinfo{author}{\bibfnamefont{I.}~\bibnamefont{Lindgren}} \bibnamefont{and}
  \bibinfo{author}{\bibfnamefont{S.}~\bibnamefont{Salomonson}},
  \bibinfo{journal}{Phys. Rev. A} \textbf{\bibinfo{volume}{67}},
  \bibinfo{pages}{056501} (\bibinfo{year}{2003}{\natexlab{a}}).

\bibitem[{\citenamefont{Lindgren and Salomonson}(2003{\natexlab{b}})}]{LS03b}
\bibinfo{author}{\bibfnamefont{I.}~\bibnamefont{Lindgren}} \bibnamefont{and}
  \bibinfo{author}{\bibfnamefont{S.}~\bibnamefont{Salomonson}},
  \bibinfo{journal}{Adv. Quantum Chem.} \textbf{\bibinfo{volume}{43}},
  \bibinfo{pages}{95} (\bibinfo{year}{2003}{\natexlab{b}}).

\bibitem[{\citenamefont{Lindgren and Salomonson}(2004)}]{LS04}
\bibinfo{author}{\bibfnamefont{I.}~\bibnamefont{Lindgren}} \bibnamefont{and}
  \bibinfo{author}{\bibfnamefont{S.}~\bibnamefont{Salomonson}},
  \bibinfo{journal}{Phys. Rev. A} p. \bibinfo{pages}{(accepted)}
  (\bibinfo{year}{2004}).

\bibitem[{\citenamefont{Nesbet}(1998)}]{Ne98}
\bibinfo{author}{\bibfnamefont{R.~K.} \bibnamefont{Nesbet}},
  \bibinfo{journal}{Phys. Rev. A} \textbf{\bibinfo{volume}{58}},
  \bibinfo{pages}{R12} (\bibinfo{year}{1998}).

\bibitem[{\citenamefont{Nesbet}(2001)}]{Ne01}
\bibinfo{author}{\bibfnamefont{R.~K.} \bibnamefont{Nesbet}},
  \bibinfo{journal}{Phys. Rev. A} \textbf{\bibinfo{volume}{65}},
  \bibinfo{pages}{010502(R)} (\bibinfo{year}{2001}).

\bibitem[{\citenamefont{Nesbet}(2003)}]{Ne03}
\bibinfo{author}{\bibfnamefont{R.~K.} \bibnamefont{Nesbet}},
  \bibinfo{journal}{Adv. Quantum Chem.} \textbf{\bibinfo{volume}{43}},
  \bibinfo{pages}{1} (\bibinfo{year}{2003}).

\bibitem[{\citenamefont{Lindgren and Morrison}(1986)}]{LM86}
\bibinfo{author}{\bibfnamefont{I.}~\bibnamefont{Lindgren}} \bibnamefont{and}
  \bibinfo{author}{\bibfnamefont{J.}~\bibnamefont{Morrison}},
  \emph{\bibinfo{title}{\textit{Atomic Many-Body Theory}}}
  (\bibinfo{publisher}{Second edition, Springer-Verlag},
  \bibinfo{address}{Berlin}, \bibinfo{year}{1986}).

\bibitem[{\citenamefont{Salomonson and {\"O}ster}(1989)}]{SO89a}
\bibinfo{author}{\bibfnamefont{S.}~\bibnamefont{Salomonson}} \bibnamefont{and}
  \bibinfo{author}{\bibfnamefont{P.}~\bibnamefont{{\"O}ster}},
  \bibinfo{journal}{Phys. Rev. A} \textbf{\bibinfo{volume}{40}},
  \bibinfo{pages}{5559} (\bibinfo{year}{1989}).

\bibitem[{\citenamefont{Frankowski and Pekeris}(1966)}]{FP66}
\bibinfo{author}{\bibfnamefont{K.}~\bibnamefont{Frankowski}} \bibnamefont{and}
  \bibinfo{author}{\bibfnamefont{C.~L.} \bibnamefont{Pekeris}},
  \bibinfo{journal}{Phys. Rev.} \textbf{\bibinfo{volume}{146}},
  \bibinfo{pages}{46} (\bibinfo{year}{1966}).

\bibitem[{\citenamefont{Freund et~al.}(1984)\citenamefont{Freund, Huxtable, and
  Morgan}}]{FHM84}
\bibinfo{author}{\bibfnamefont{D.~E.} \bibnamefont{Freund}},
  \bibinfo{author}{\bibfnamefont{B.~D.} \bibnamefont{Huxtable}},
  \bibnamefont{and} \bibinfo{author}{\bibfnamefont{J.~D.}
  \bibnamefont{Morgan}}, \bibinfo{journal}{Phys. Rev. A}
  \textbf{\bibinfo{volume}{29}}, \bibinfo{pages}{980} (\bibinfo{year}{1984}).

\bibitem[{\citenamefont{Koga et~al.}(1993)\citenamefont{Koga, Kasai, and
  Thakkar}}]{Koga93}
\bibinfo{author}{\bibfnamefont{T.}~\bibnamefont{Koga}},
  \bibinfo{author}{\bibfnamefont{Y.}~\bibnamefont{Kasai}}, \bibnamefont{and}
  \bibinfo{author}{\bibfnamefont{A.~J.} \bibnamefont{Thakkar}},
  \bibinfo{journal}{Int. J. Quantum Chem.} \textbf{\bibinfo{volume}{46}},
  \bibinfo{pages}{689} (\bibinfo{year}{1993}).

\bibitem[{\citenamefont{Lindgren}(1971)}]{Li71}
\bibinfo{author}{\bibfnamefont{I.}~\bibnamefont{Lindgren}},
  \bibinfo{journal}{Int. J. Quantum Chem.} \textbf{\bibinfo{volume}{5}},
  \bibinfo{pages}{411} (\bibinfo{year}{1971}).

\bibitem[{\citenamefont{Talman and Shadwick}(1976)}]{TS76}
\bibinfo{author}{\bibfnamefont{J.~D.} \bibnamefont{Talman}} \bibnamefont{and}
  \bibinfo{author}{\bibfnamefont{W.~S.} \bibnamefont{Shadwick}},
  \bibinfo{journal}{Phys. Rev. A} \textbf{\bibinfo{volume}{14}},
  \bibinfo{pages}{36} (\bibinfo{year}{1976}).

\bibitem[{\citenamefont{Perdew and Zunger}(1981)}]{PZ81}
\bibinfo{author}{\bibfnamefont{J.~P.} \bibnamefont{Perdew}} \bibnamefont{and}
  \bibinfo{author}{\bibfnamefont{A.}~\bibnamefont{Zunger}},
  \bibinfo{journal}{Phys. Rev. B} \textbf{\bibinfo{volume}{23}},
  \bibinfo{pages}{5048} (\bibinfo{year}{1981}).

\bibitem[{\citenamefont{Pekeris}(1959)}]{Pek59}
\bibinfo{author}{\bibfnamefont{C.~L.} \bibnamefont{Pekeris}},
  \bibinfo{journal}{Phys. Rev.} \textbf{\bibinfo{volume}{115}},
  \bibinfo{pages}{1216} (\bibinfo{year}{1959}).

\end{thebibliography}
\end{document}